\pdfoutput=1
\documentclass[11pt]{article}
\usepackage[pdftex]{graphicx,color} 
\usepackage{jheppub}
\usepackage{amsmath}
\usepackage{amssymb}

\usepackage{comment}
\usepackage{multirow}

\newcommand{\Tr}{\textrm{Tr}}

\renewcommand{\a}{\alpha}
\renewcommand{\b}{\beta}

\newcommand\dd{\mathrm{d}}
\newcommand{\bali}{\begin{align}}
\newcommand{\eali}{\end{align}}
\newcommand{\bea}{\begin{equation}\begin{aligned}}
\newcommand{\eea}[1]{\label{#1}\end{aligned}\end{equation}}
\newcommand{\beg}{\begin{equation}\begin{gathered}}
\newcommand{\eeg}[1]{\label{#1}\end{gathered}\end{equation}}

\newcommand{\bt}[1]{{\mathord{\vcenter{\hbox{\includegraphics[scale=0.4]{{{#1}}}}}}}}
\newcommand{\btm}[1]{{\mathord{\vcenter{\hbox{\includegraphics[scale=0.3]{{{#1}}}}}}}}
\newcommand{\bts}[1]{{\mathord{\vcenter{\hbox{\includegraphics[scale=0.15]{{{#1}}}}}}}}

\graphicspath{ {./figures/} }

\newcommand{\beq}{\begin{equation}}
\newcommand{\eeq}{\end{equation}}

\newcommand{\bra}{\big \langle}
\newcommand{\ket}{\big \rangle}
\newcommand{\nz}{n_{Z}}
\newcommand{\ntheta}{n_{\Theta}}

\newcommand{\calO}{\mathcal{O}}
\newcommand{\calN}{\mathcal{N}}
\newcommand{\calS}{\mathcal{S}}
\newcommand{\bz}{{\bf z}}
\newcommand{\btheta}{{\bf \theta}}
\newcommand{\bZ}{{\bf Z}}
\newcommand{\bTheta}{{\bf \Theta}}
\newcommand{\bD}{{\bf D}}
\newcommand{\bbeta}{\boldsymbol{\beta}}
\newcommand{\bgamma}{\boldsymbol{\gamma}}
\newcommand{\De}{\Delta}
\newcommand{\G}{\Gamma}

\newlength\Colsep
\title{Conformal correlators of mixed-symmetry tensors}
\author{Miguel S. Costa$ ^{\dagger,\ddagger}$}
\author{and Tobias Hansen$ ^\sharp$}
\affiliation{$ ^\dagger$ Centro de F\'\i sica do Porto,
Departamento de F\'\i sica e Astronomia\\
Faculdade de Ci\^encias da Universidade do Porto\\
Rua do Campo Alegre 687,
4169--007 Porto, Portugal}
\affiliation{$ ^\ddagger$ Theory Division, Department of Physics, CERN\\
CH-1211 Gen\`eve 23, Switzerland }
\affiliation{$ ^\sharp$ II. Institut f\"ur Theoretische Physik, Universit\"at Hamburg\\ Luruper Chaussee 149, D-22761 Hamburg, Germany }
\emailAdd{miguelc@fc.up.pt, tobias.hansen@desy.de}

\preprint{CERN-PH-TH-2015-003}


\abstract{We generalize the embedding formalism for conformal field theories to the case of  general operators with mixed symmetry.
The index-free notation encoding symmetric tensors as polynomials in an auxiliary polarization vector is extended to mixed-symmetry tensors by introducing a
new commuting or anticommuting polarization vector for each row or column in the Young diagram that describes the index
symmetries of the tensor.
We determine the tensor structures that are allowed in $n$-point conformal correlation functions and give an algorithm for counting them in terms of
tensor product coefficients. 
A simple derivation of the  unitarity bound for arbitrary mixed-symmetry tensors
is obtained by considering the conservation condition in embedding space.
We show, with an example, how the new formalism can be used to compute 
conformal blocks of arbitrary external fields for the exchange of any conformal primary and its descendants.
The matching between the number of tensor structures in conformal field theory correlators of operators 
in $d$ dimensions and massive scattering amplitudes in $d+1$ dimensions is also seen to carry over to mixed-symmetry tensors. }

\begin{document}
\maketitle

\section{Introduction}

The study of Conformal Field Theories (CFTs) is among the most important subjects in theoretical physics, with implications to critical phenomena, particle physics and, in the light of the AdS/CFT duality,  quantum gravity. In past years we have witnessed a revival in the study of CFTs in dimensions higher than two. This study is considerably more difficult than the two-dimensional case, where  the conformal group possesses an infinitely dimensional extension given by the Virasoro algebra, which leads to many known exactly solvable models.
On one hand, the {\em conformal bootstrap} program \cite{Ferrara:1973yt,Polyakov:1974gs} applied to higher dimensional CFTs, revived in \cite{Rattazzi:2008pe}, has already shown its merits by providing the most accurate computation to date of  3D Ising model critical exponents \cite{ElShowk:2012ht,El-Showk:2014dwa,Kos:2014bka}. On the other hand, the most studied case of the AdS/CFT duality \cite{Maldacena:1997re} considers ${\cal N}=4$ Super Yang-Mills, which is a four-dimensional CFT. In particular, this theory is believed to be integrable in the planar limit \cite{Minahan:2002ve,Bena:2003wd}, thus providing the first example of an exactly solvable 4D gauge theory.

To advance in the {\em conformal bootstrap} program, as well as our understanding of  AdS/CFT, it is necessary to further develop analytic and computational techniques to deal with arbitrary tensor primary fields. In $d$  dimensions, these fields are classified by the unitary irreducible representations of the conformal group $SO(d+1,1)$, which are labeled by the conformal dimension $\Delta$ and by an irreducible representation (irrep) of $SO(d)$. A first step in this direction was made in \cite{Costa:2011mg}, where symmetric tensors of arbitrary spin were studied in detail. The goal of this paper is to extend this work by considering $SO(d)$ tensors with mixed symmetry. 

We shall start, in section two, with the general classification of irreducible tensor representations of  $SO(d)$, which can be represented by Young diagrams.
This is a well known subject, which we shall review in order to introduce the reader to the necessary formalism. We will then see how to encode, in general,
mixed-symmetry tensors in terms of polynomials of polarization vectors. To encode their mixed symmetry it is necessary to employ a combination of Grassmann valued and ordinary commuting polarizations. Actual computations simplify considerably if fields that live in 
$d$-dimensional Euclidean space  are embedded in an auxiliary $(d+2)$-dimensional Minkowski space, where the conformal group $SO(d+1,1)$ 
acts linearly as the usual Lorentz transformations. We shall see that this formalism can be easily extended to include mixed-symmetry tensors by encoding them in
polynomials of polarization vectors in the embedding space.

In section three we show how to construct CFT $n$-point correlation functions of arbitrary tensors. The formalism is presented in general terms, but we shall give a number of simple examples for  two-, three- and four-point functions,
so the reader can appreciate the  simplicity and efficiency of the method.
We will also describe the general case of $n$-point functions.

In section four we consider arbitrary conserved tensors. We see how to implement the conservation equation in the embedding formalism, and also how to derive the unitary bound for 
conserved tensors in arbitrary irreducible $SO(d)$ representations.

As an application of the new formalism we consider, in section five, the problem of computing conformal blocks for any desired external primary fields, describing the exchange  
of an arbitrary conformal primary and its descendants. With the help of shadow operators, these conformal blocks can be written as an integral of
three-point functions, leading to an expression of the conformal blocks in terms of a finite number of integrals, which can be expressed in terms of
hypergeometric functions for even dimensions \cite{SimmonsDuffin:2012uy}. To see the method at work, we shall consider explicitly the example of the four-point function of two scalars and two vectors, 
exchanging a mixed-symmetry tensor of rank three. 

In section six we show that the number of tensor structures in CFT correlators of non-conserved mixed-symmetry tensors
in $d$ dimensions matches that of massive scattering amplitudes in $d+1$ dimensions, as expected. 
Section seven presents final comments.

\section{Mixed-symmetry tensors}

\subsection{Parametrizing Young diagrams}

In this paper the traceless irreducible tensor representations of $SO(d)$ are considered.
These representations are enumerated mostly\footnote{
There is a one-to-one correspondence between traceless irreducible tensor representations of $SO(d)$
and the Young diagrams satisfying \eqref{eq:irrep_height_restriction}
except for the case $d=2n, h^{\lambda}_1 = n$  \cite{MR885807}. In this case the representation
with the symmetry corresponding to $\lambda$ can be decomposed further using the
Levi-Civita tensor and is therefore not irreducible.
A well-known example is the decomposition of the two-form in four dimensions
into self-dual and anti-self-dual parts.
}
by Young diagrams, which encode the (anti-) symmetry of the tensors
under permutation of their indices.

There are two different ways to parametrize the shape of a Young diagram $\lambda$.
The first is by giving a partition $l^\lambda = \left(l^\lambda_1, l^\lambda_2, \ldots\right)$ containing the
lengths of the rows, $l^\lambda_i$ being the length of the $i$-th row.
The diagram that is obtained from $\lambda$ by exchanging rows and columns is called the transpose $\lambda^t$.
The partition $h^{\lambda}$ describes the column heights of $\lambda$ and is the conjugate partition
to $l^\lambda$, $l^{\lambda^t} \equiv h^\lambda = \left(h^\lambda_1, h^\lambda_2, \ldots\right)$.
A second way to describe the shape of a Young diagram is by its Dynkin label $\lambda = \big[\lambda_1,\lambda_2,\ldots,\lambda_{h^{\lambda}_1}\big]$,
which lists the numbers $\lambda_i$ of columns with $i$ boxes.
Apart from the exception mentioned in the footnote,
the Young diagram $\lambda$ labels an irrep of $SO(d)$ if and only if its overall height $h^{\lambda}_1$
does not exceed the rank of the Lie algebra corresponding to $SO(d)$,
\beq
h^{\lambda}_1 \leq \Big\lfloor \frac{d}{2} \Big\rfloor 
=
\begin{cases}\frac{d}{2}\,, \quad &d \text{ even}\,, \\
\frac{d-1}{2}\,, \quad &d \text{ odd}\,.
\end{cases}
\label{eq:irrep_height_restriction}
\eeq
The total number of boxes is denoted by $|\lambda|$,
\beq
|\lambda| = \sum\limits_{i} i \lambda_i = \sum\limits_{i} l^\lambda_i = \sum\limits_{i} h^\lambda_i\,.
\eeq
It will be useful to label the number of rows with more than one box
$\nz^\lambda$ and the number of columns with more than
one box $\ntheta^{\lambda}$,
\beq
\nz^\lambda = \sum\limits_{i=2}^{l^{\lambda}_1} \lambda^t_i \,, \qquad
\ntheta^\lambda = \sum\limits_{i=2}^{h^{\lambda}_1} \lambda_i \,.
\eeq
All of this is best illustrated by the following example:

\noindent\begin{minipage}{\textwidth}
\begin{minipage}[c][3cm][c]{\dimexpr0.25\textwidth-0.5\Colsep\relax}
\begin{equation*}
\hspace{30pt}\bt{young_2102}\hspace{-40pt}
\end{equation*}
\end{minipage}\hfill
\begin{minipage}[c][3cm][c]{\dimexpr0.75\textwidth-0.5\Colsep\relax}
\bea
\lambda &= [2,1,0,2] \,, \qquad \quad &|\lambda|&=12\,,\\
l^{\lambda} &= (5,3,2,2) \,, \qquad &\nz^\lambda &=4\,,\\
h^{\lambda} &= (4,4,2,1,1) \,, \qquad &\ntheta^\lambda &=3\,.
\eea{eq:young_example}
\end{minipage}
\end{minipage}
The $\lambda$ on $l^\lambda$, $h^\lambda$, $\nz^{\lambda}$ and $\ntheta^{\lambda}$
will frequently be omitted or replaced by $i$ if the Young diagram is of shape $\lambda_i$.

\subsection{Birdtracks and Grassmann variables}
\label{sec:birdtracks}

Probably the best way to think about mixed-symmetry tensors is in terms of birdtrack notation\footnote{
See \cite{Cvitanovic:2008zz} for a beautiful group theory book entirely in terms of birdtracks.}
where index contractions are simply drawn as lines
\beq
\btm{2lines} = \delta^{a_1 b_1} \delta^{a_2 b_2} \,.
\eeq
Symmetrization and antisymmetrization are indicated by the symbols
\bea
\btm{def_sym} &= \frac{1}{n!} \left\{ \btm{def_0twists} + \btm{def_1twists} + \btm{def_2twists} + \ldots \right\} ,\\
\btm{def_asym} &= \frac{1}{n!} \left\{ \btm{def_0twists} - \btm{def_1twists} + \btm{def_2twists} - \ldots \right\} .
\eea{eq:def_bt_sym_asym}
This notation has the advantage that it makes it immediately visible when terms are vanishing because two or more symmetric indices are antisymmetrized or vice versa
\beq
\btm{bt_overlap} = 0\,.
\eeq
Furthermore, birdtracks can be diagrammatically transformed, for example using that repeated (anti)symmetrizations of subsets of indices have no effect
\beq
\btm{bt_2xsym} = \btm{bt_1xsym} \ , \quad \qquad \btm{bt_2xasym} = \btm{bt_1xasym} \ .
\eeq

A symmetrized contraction of $n$ indices is generated by the $n$-th derivative of $n$ components of
an auxiliary vector $z$,
\beq
\btm{def_sym} = \frac{1}{n!}\, \partial_{z_{a_1}} \ldots \partial_{z_{a_n}} z_{b_1} \ldots z_{b_n} \,.
\eeq
Antisymmetrization works analogously with an auxiliary vector in Grassmann variables $\theta$,
\beq
\btm{def_asym} = \frac{1}{n!} \,\partial_{\theta_{a_1}} \ldots \partial_{\theta_{a_n}} \theta_{b_1} \ldots \theta_{b_n} \,.
\label{eq:asym_derivatives}
\eeq
The Grassmann variables are anticommuting in the sense that
\beq
\theta^{(p)}_{a} \theta^{(q)}_{b} = (-1)^{\delta^{pq}} \theta^{(q)}_{b} \theta^{(p)}_{a} \,.
\eeq
Here an additional label $(p)$ was introduced to allow for several independent antisymmetrizations at the same time.
Derivatives with respect to Grassmann variables are implied to be right derivatives,
\beq
\partial_{\theta^{(r)}_c} \theta^{(p)}_{a} \theta^{(q)}_{b}
= \delta^{rq}\delta^{cb} \theta^{(p)}_{a} + (-1)^{\delta^{pq}} \delta^{rp}\delta^{ca} \theta^{(q)}_{b}\,.
\eeq

\subsection{Young symmetrization and antisymmetric basis}
\label{sec:mixed-symmetry_tensors}

The symmetry of a Young diagram is imposed on a tensor via Young symmetrizers. Each row of the
diagram corresponds to a symmetrization and each column corresponds to an antisymmetrization.
This can be nicely illustrated by an example, following \cite{Cvitanovic:2008zz}.
To actually write down components of mixed-symmetry tensors it is necessary to choose a basis for the 
irreducible representation at hand. This requires an assignment between the boxes of the Young
diagram and the indices of the tensor. Therefore the bases of the irreps under consideration
are labeled by Young tableaux.
A symmetrizer given by the Young tableau $\mathbf{YT}$ creates the tensor $T^\mathbf{YT}$
with appropriate symmetry from a generic tensor $T$, 
\beq
\mathbf{YT} = \bt{youngt_211s} \qquad \rightarrow \qquad T^\mathbf{YT} = {\mathord{\vcenter{\hbox{\scalebox{0.3}{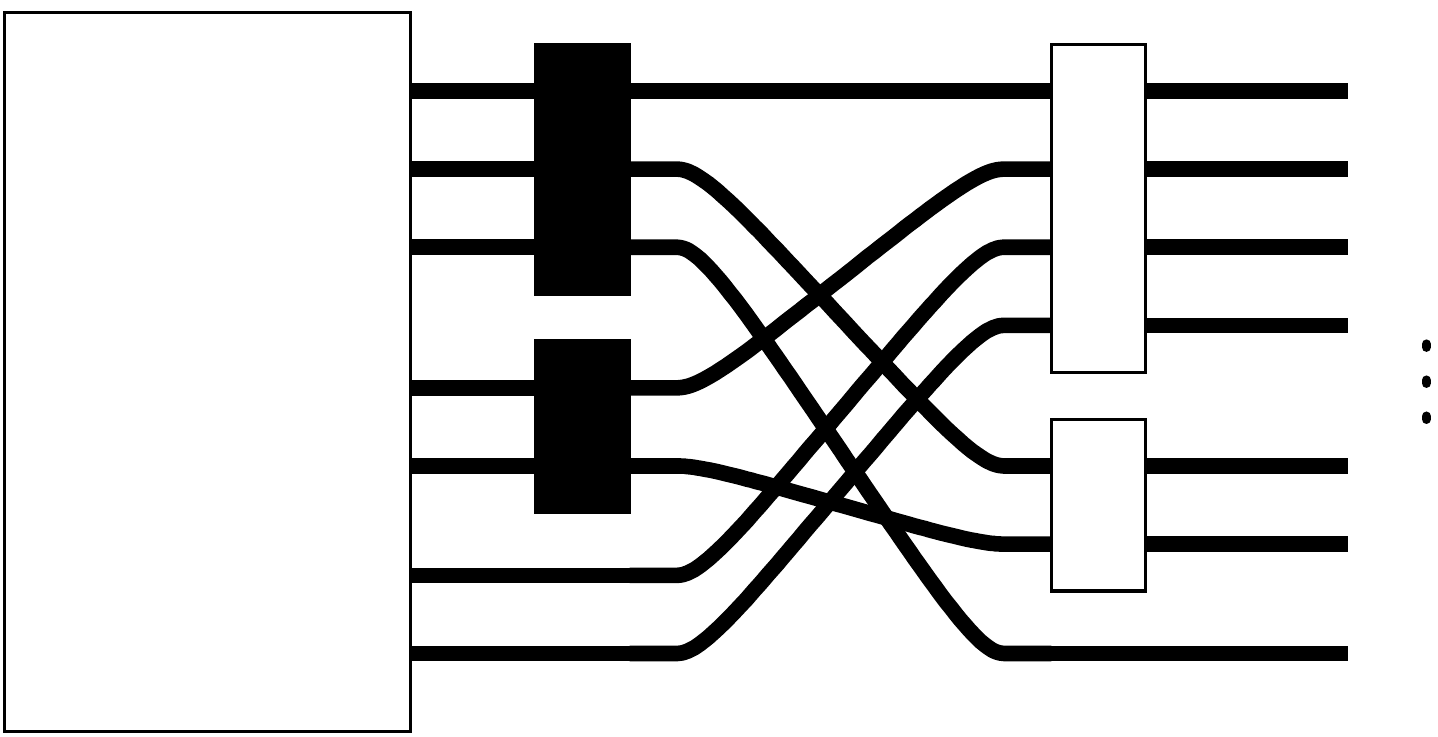}}}}}\quad .
\label{eq:sym_tensor}
\eeq
This tensor has the manifest symmetry properties
\beq
T^\mathbf{YT}_{a_1 a_2 a_3 a_4 a_5 a_6 a_7} = T^\mathbf{YT}_{(a_1 a_2 a_3 a_4)(a_5 a_6) a_7} \,,
\eeq
but there are also less obvious symmetries caused by the antisymmetrizations. Due to the manifest symmetries,
$T^\mathbf{YT}$ is said to belong to the symmetric basis. The antisymmetric basis is obtained by changing the 
order of symmetrization and antisymmetrization
\beq
\mathbf{YT}' = \bt{youngt_211a} \qquad \rightarrow \qquad T^{\mathbf{YT}'} = {\mathord{\vcenter{\hbox{\scalebox{0.3}{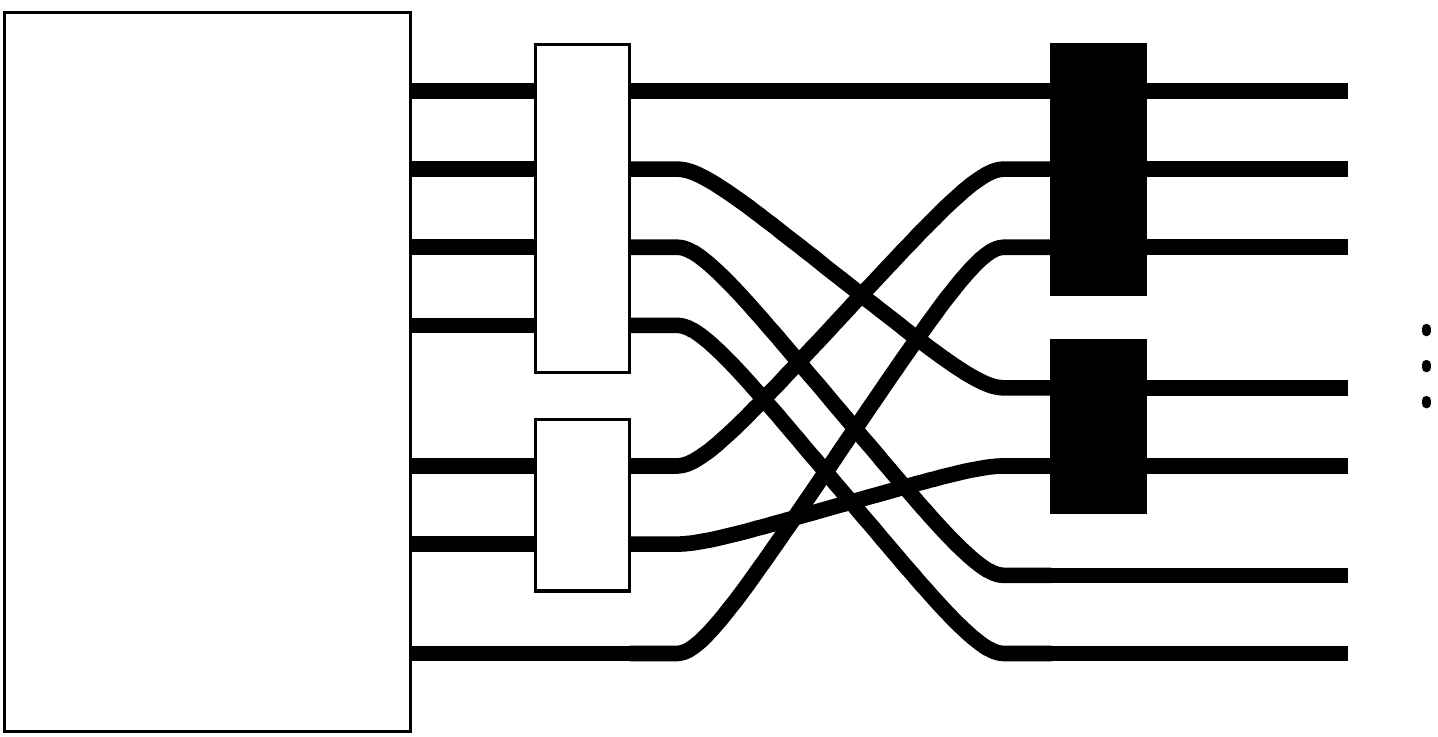}}}}}\quad .
\label{eq:asym_tensor}
\eeq
Here we have manifest antisymmetry
\beq
T^{\mathbf{YT}'}_{a_1 a_2 a_3 a_4 a_5 a_6 a_7} = T^{\mathbf{YT}'}_{[a_1 a_2 a_3][a_4 a_5](a_6 a_7)} \,.
\eeq
The only reason we used a different Young tableau for this second example is to spare us from having to cross lines on the
right hand side of the birdtrack diagram. 
We will in this paper work only in antisymmetric bases with Young
tableaux where the boxes are enumerated column by column, as in \eqref{eq:asym_tensor}.
The tensors corresponding to different bases (different tableaux)
can be obtained simply by commutation of indices.

It may also be instructive to see how the non-explicit index symmetries
manifest themselves on the components of the tensors,
again in the antisymmetric basis with boxes labeled column by column.
To this end assign different labels to each anticommuting group of indices
\beq
f_{a_1\ldots a_{h_1}b_1\ldots b_{h_2}c_1\ldots c_{h_3}\ldots g_1\ldots g_{h_{l_1}}}
= f_{[a_1\ldots a_{h_1}][b_1\ldots b_{h_2}][c_1\ldots c_{h_3}]\ldots[g_1\ldots g_{h_{l_1}}]} \,.
\eeq
Apart from the antisymmetry, the Young symmetrization implies that the antisymmetrization of any of the indices $b$ with all the $a$
vanishes, as well as the antisymmetrization of any of the $c$ with all indices $a$ or all $b$ and so forth \cite{MR1464693}.
Explicitly this means that
\begin{align}
& f_{[a_1\ldots a_{h_1}][b_1\ldots b_{h_2}][c_1\ldots c_{h_3}]\ldots [g_1\ldots g_{h_{l_1}}]}
\label{eq:antisymmetrization_ex}\\
={}& f_{[b_1 a_2\ldots a_{h_1}][a_1 b_2\ldots b_{h_2}] [c_1\ldots}
+ f_{[a_1 b_1 a_3\ldots a_{h_1}][a_2 b_2\ldots b_{h_2}][c_1\ldots}
+ \ldots + f_{[a_1\ldots a_{h_1-1}b_1 ][a_{h_1} b_2\ldots b_{h_2}][c_1\ldots} 
\nonumber\\
={}& f_{[c_1 a_2\ldots a_{h_1}][ b_1\ldots b_{h_2}][ a_1 c_2\ldots }
+ f_{[a_1 c_1 a_3\ldots a_{h_1} ][b_1\ldots b_{h_2} ][a_2 c_2\ldots }
+ \ldots + f_{[a_1\ldots a_{h_1-1}c_1 ][b_1\ldots b_{h_2} ][a_{h_1} c_2\ldots } \,.
\nonumber
\end{align}
There are also more general relations that arise from exchanging $k$ indices from one column
with all possible $k$-element subsets of a column to its left. Here the order of the two sets of
indices is kept, so that the right hand side of the general equation has $\binom{h_l}{k}$ terms if the
left column has height $h_l$.
As a special case of these relations the tensors are symmetric under exchange
of complete groups of antisymmetric indices
if the corresponding columns in the Young tableau are of equal height, e.g.\ for $h_2 = h_3$,
\beq
f_{[a_1\ldots a_{h_1}][b_1\ldots b_{h_2}][c_1\ldots c_{h_3}]\ldots [g_1\ldots g_{h_{l_1}}]}
= f_{[a_1\ldots a_{h_1}][c_1\ldots c_{h_3}][ b_1\ldots b_{h_2}] \ldots [g_1\ldots g_{h_{l_1}}]} \,.
\label{eq:index_groups_sym}
\eeq
Since it will be needed in Section \ref{sec:conserved_tensors} we also state the equation analogous to
\eqref{eq:antisymmetrization_ex}
for a tensor
\beq
f_{a_1\ldots a_{l_1}b_1\ldots b_{l_2}c_1\ldots c_{l_3}\ldots g_1\ldots g_{l_{h_1}}}
= f_{(a_1\ldots a_{l_1})(b_1\ldots b_{l_2})(c_1\ldots c_{l_3})\ldots(g_1\ldots g_{l_{h_1}})} \,,
\label{sym_basis_explicit}
\eeq
in the symmetric basis with boxes enumerated row by row as in \eqref{eq:sym_tensor},
\begin{align}
& -f_{(a_1\ldots a_{l_1})(b_1\ldots b_{l_2})(c_1\ldots c_{l_3})\ldots (g_1\ldots g_{l_{h_1}})}
\label{eq:symmetrization_ex}\\
={}& f_{(b_1 a_2\ldots a_{l_1})(a_1 b_2\ldots b_{l_2}) (c_1\ldots}
+ f_{(a_1 b_1 a_3\ldots a_{l_1})(a_2 b_2\ldots b_{l_2})(c_1\ldots}
+ \ldots + f_{(a_1\ldots a_{l_1-1}b_1 )(a_{l_1} b_2\ldots b_{l_2})(c_1\ldots} 
\nonumber\\
={}& f_{(c_1 a_2\ldots a_{l_1})( b_1\ldots b_{l_2})( a_1 c_2\ldots }
+ f_{(a_1 c_1 a_3\ldots a_{l_1} )(b_1\ldots b_{l_2} )(a_2 c_2\ldots }
+ \ldots + f_{(a_1\ldots a_{l_1-1}c_1 )(b_1\ldots b_{l_2} )(a_{l_1} c_2\ldots } \,.
\nonumber
\end{align}
\subsection{Encoding mixed-symmetry tensors by polynomials}
\label{sec:index_free_notation}

In general, to encode a mixed-symmetry tensor by a polynomial, the strategy is to contract it with a tensor
with the same mixed symmetry, which is built out of auxiliary polarizations. 
To construct a Young symmetrized tensor in the antisymmetric basis out of auxiliary vectors, one can start 
with a set of polarizations that is already symmetrized so that only the antisymmetrization is left to do.
For the example \eqref{eq:asym_tensor}, the following tensor depending on the auxiliary vectors
$z^{(1)}$, $z^{(2)}$ and $z^{(3)}$ is appropriately symmetrized
\beq
{\mathord{\vcenter{\hbox{\scalebox{0.3}{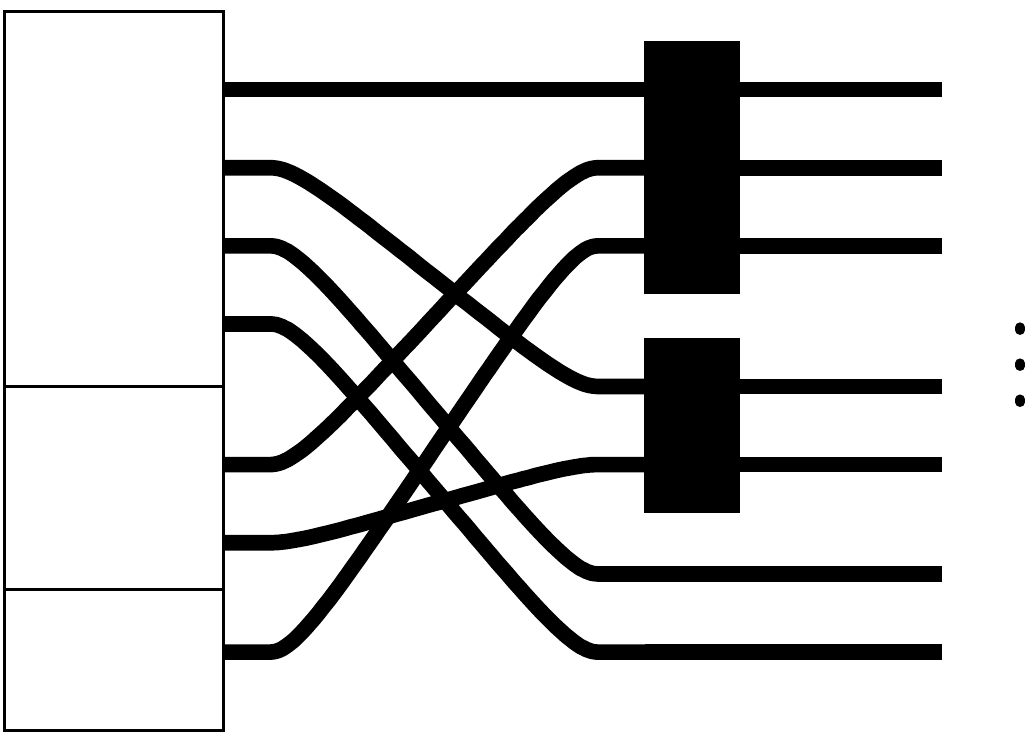}}}}}
\quad .
\label{eq:asym_poly_ex}
\eeq
Using \eqref{eq:asym_derivatives} to encode the antisymmetrization, \eqref{eq:asym_poly_ex} can be written as
\begin{align}
&\frac{1}{3!2!}\left( z^{(1)} \cdot \partial_{\theta^{(1)}} \right)
\left( z^{(1)} \cdot \partial_{\theta^{(2)}} \right)
\left( z^{(1)} \cdot \partial_{\theta^{(3)}} \right)
\left( z^{(1)} \cdot \partial_{\theta^{(4)}} \right)
\nonumber\\
&\left( z^{(2)} \cdot \partial_{\theta^{(1)}} \right)
\left( z^{(2)} \cdot \partial_{\theta^{(2)}} \right)
\left( z^{(3)} \cdot \partial_{\theta^{(1)}} \right)
\theta^{(1)}_{a_1}
\theta^{(1)}_{a_2}
\theta^{(1)}_{a_3}
\theta^{(2)}_{a_4}
\theta^{(2)}_{a_5}
\theta^{(3)}_{a_6}
\theta^{(4)}_{a_7} \,.
\end{align}
This can be shortened by avoiding the introduction of polarizations
that appear only once and hence do not cause any (anti-)symmetrization,
i.e.\ doing  explicitly the derivatives in the polarizations $\theta^{(3)}$ and $\theta^{(4)}$,
\beq
\left( z^{(1)} \cdot \partial_{\theta^{(3)}} \right)
\left( z^{(1)} \cdot \partial_{\theta^{(4)}} \right)
\theta^{(3)}_{a_6}
\theta^{(4)}_{a_7}
=
z^{(1)}_{a_6}
z^{(1)}_{a_7} \,.
\eeq
After this step the symmetry in the indices $a_6$ and $a_7$
is manifest. Likewise, $z^{(3)}$ that appears only once in this example
through the derivative $\left(z^{(3)} \cdot \partial_{\theta^{(1)}}\right)$,
does not encode any symmetry. More generally, for diagrams with more than one row of length one,
the action of such derivatives hides antisymmetry. We shall therefore omit these derivative terms, with the result that
the encoding polynomial will depend not only on symmetric polarizations,
but also on $\theta^{(1)}$, therefore making antisymmetrization explicit on the indices corresponding to all rows of length one.

Thus, the slightly less elegant, but more pragmatic Young symmetric
polarization we  use  for the example at hand will be the polynomial in ${\bf z}\equiv \left(z^{(1)},z^{(2)},\theta^{(1)}  \right)$ given by
\beq
\left( z^{(1)} \cdot \partial_{\theta^{(1)}} \right)
\left( z^{(1)} \cdot \partial_{\theta^{(2)}} \right)
\left( z^{(2)} \cdot \partial_{\theta^{(1)}} \right)
\left( z^{(2)} \cdot \partial_{\theta^{(2)}} \right)
\theta^{(1)}_{a_1}
\theta^{(1)}_{a_2}
\theta^{(1)}_{a_3}
\theta^{(2)}_{a_4}
\theta^{(2)}_{a_5}
z^{(1)}_{a_6}
z^{(1)}_{a_7} \,,
\eeq
which is  quartic in $z^{(1)}$, quadratic in $z^{(2)}$ and linear in $\theta^{(1)}$, as appropriate for a Young diagram with
lengths of rows given by
$l^\lambda=(4,2,1)$. This Young symmetric
polarization is obtained by acting with derivatives of the type $\left(z^{(p)} \cdot \partial_{\theta^{(q)}}\right)$ on a polynomial
in ${\boldsymbol \theta}\equiv\left(\theta^{(1)},\theta^{(2)},z^{(1)}\right)$,  cubic  in $\theta^{(1)}$, quadratic in $\theta^{(2)}$ and quadratic in $z^{(1)}$, as appropriate for a Young diagram with
lengths of columns given by
$h^\lambda=(3,2,1,1)$. A tensor with components $f^{a_1\ldots a_7}$ in the irrep of this example will then be encoded  by the polynomial
\bea
f({\bf z})
\equiv{}
\left( z^{(1)} \cdot \partial_{\theta^{(1)}} \right)
\left( z^{(1)} \cdot \partial_{\theta^{(2)}} \right)
\left( z^{(2)} \cdot \partial_{\theta^{(1)}} \right)
\left( z^{(2)} \cdot \partial_{\theta^{(2)}} \right)
\bar f( {\boldsymbol \theta})  \,,
\eea{eq:Example_poly_asym}
where
\bea
\bar f({\boldsymbol \theta}) 
\equiv{}&
\theta^{(1)}_{a_1}
\theta^{(1)}_{a_2}
\theta^{(1)}_{a_3}
\theta^{(2)}_{a_4}
\theta^{(2)}_{a_5}
z^{(1)}_{a_6}
z^{(1)}_{a_7} 
f^{a_1\ldots a_7} \,.
\eea{eq:fbar_example}
Notice that, in this example, the assignment of  the polarization vectors in ${\bf z}$ and in ${\boldsymbol \theta}$ 
to the boxes of the Young diagram is done according to
\beq
{\mathord{\vcenter{\hbox{\scalebox{0.4}{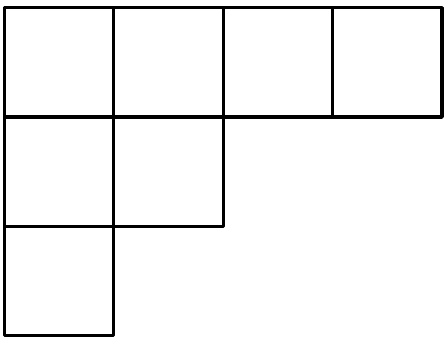}}}}}\quad\quad{\rm and}\quad\quad\quad{\mathord{\vcenter{\hbox{\scalebox{0.4}{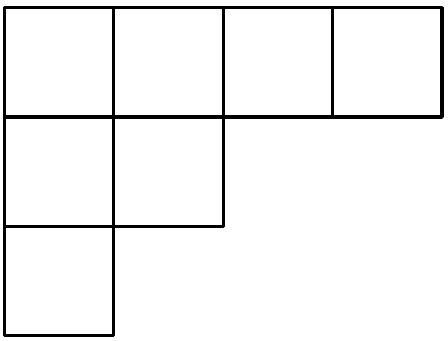}}}}}\quad,
\eeq
respectively.

In general we shall consider  $\ntheta$ anticommuting and $\nz$ commuting polarization vectors for a given tensor operator.
A convenient notation for the mostly anticommuting polarizations which
are first contracted to the tensor is
\beq
{\boldsymbol \theta} \equiv \left( \theta^{(1)},\theta^{(2)}.\ldots,\theta^{(\ntheta)},z^{(1)} \right).
\eeq
In cases where there are no columns with one box the last entry is absent and $\boldsymbol \theta$ contains only anti-commuting 
polarizations.
Similarly, we will write for the mostly commuting polarizations on which the final encoding
polynomial depends
\beq
{\bf z} \equiv \left(z^{(1)},z^{(2)},\ldots,z^{(\nz)},\theta^{(1)}\right).
\eeq
Again, in cases where there are no rows with one box the last entry is absent and ${\bf z}$ contains only commuting polarizations.
Generalizing the previous example, we have that  a 
tensor $f^{a_1\ldots a_{|\lambda|}}$ in the irrep $\lambda$ is encoded by the polynomial
\bea
f({\bf z})
\equiv{}
\prod\limits_{p=1}^{\nz} \prod\limits_{q=1}^{\min ( l_p,\ntheta)}
\left( z^{(p)} \cdot \partial_{\theta^{(q)}} \right)
\bar f({\boldsymbol \theta})  \,,
\eea{eq:def_poly_asym}
where
\bea
\bar f({\boldsymbol \theta}) 
\equiv{}&
\theta^{(1)}_{a_1} \ldots \theta^{(1)}_{a_{h_1}} \theta^{(2)}_{a_{h_1+1}} \ldots \theta^{(2)}_{a_{h_1+h_2}} \\
&\ldots
\theta^{(\ntheta)}_{a_{h_1+\ldots+h_{\ntheta-1}+1}} \ldots  \theta^{(\ntheta)}_{a_{h_1+\ldots+h_{\ntheta}}} 
z^{(1)}_{a_{|\lambda|-\lambda_1+1}} \ldots z^{(1)}_{a_{|\lambda|}}
f^{a_1\ldots a_{|\lambda|}} \,.
\eea{eq:fbar_def}
When there are more than one row with one box, the dependence of $f({\bf z})$ on $\theta^{(1)}$ makes manifest the antisymmetry of the indices corresponding to such boxes. Likewise,
when there are more than one column with one box, the dependence of $\bar f({\boldsymbol \theta})$ on $z^{(1)}$ makes manifest the symmetry of the indices corresponding to such boxes.

The condition that $f^{a_1\ldots a_{|\lambda|}}$ is traceless can be 
used to choose the polarizations to have vanishing products
\bea
f^{a_1\ldots a_{|\lambda|}} \text{ traceless } &\leftrightarrow
\left.\bar f({\boldsymbol \theta}) \right|_{\theta^{(p)}\cdot\theta^{(q)} = \theta^{(p)}\cdot z^{(1)} = {z^{(1)}}^2 = 0} \,,\\
&\leftrightarrow
\left.f\left({\bf z}\right) \right|_{z^{(p)}\cdot z^{(q)} = z^{(p)}\cdot \theta^{(1)} = 0} \,.
\eea{eq:traceless}
This means that all terms in the tensor proportional to Kronecker deltas $\delta^{a_i a_j}$
are discarded. They have to be restored by projection to traceless tensors if one wishes to extract the tensor
from the polynomial.

To extract the tensor $f^{a_1\ldots a_{|\lambda|}}$ back from the polynomials one 
can simply restore the indices by acting with $|\lambda|$ derivatives on the polarizations
and then project to the irreducible representation $\lambda$ with the projector 
$\pi_{\lambda}^{a_1\ldots a_{|\lambda|},b_1\ldots b_{|\lambda|}}$,
\begin{align}
f^{a_1\ldots a_{|\lambda|}} ={}& \pi_{\lambda}^{a_1\ldots a_{|\lambda|},b_1\ldots b_{|\lambda|}}
 \frac{1}{h_1!}\, \partial_{\theta^{(1)}_{b_1}} \ldots \partial_{\theta^{(1)}_{b_{h_1}}} \frac{1}{h_2!}\,\partial_{\theta^{(2)}_{b_{h_1+1}}} \ldots \partial_{\theta^{(2)}_{b_{h_1+h_2}}} 
\label{eq:fbar_extract}\\
&\ldots
\frac{1}{h_{\ntheta}!} \,\partial_{\theta^{(\ntheta)}_{b_{h_1+\ldots+h_{\ntheta-1}+1}}} \ldots  \partial_{\theta^{(\ntheta)}_{b_{h_1+\ldots+h_{\ntheta}}}} 
\frac{1}{\lambda_1 !}\,\partial_{z^{(1)}_{b_{|\lambda|-\lambda_1+1}}} \ldots \partial_{z^{(1)}_{b_{|\lambda|}}}
\bar f({\boldsymbol \theta})  
\nonumber \\
={}& \pi_{\lambda}^{a_1\ldots a_{|\lambda|},b_1\ldots b_{|\lambda|}}
 \frac{1}{H(\lambda)}\,
\partial_{z^{(1)}_{b_1}} \ldots \partial_{z^{(1)}_{b_{l_1}}} 
\partial_{z^{(1)}_{b_{l_1+1}}} \ldots \partial_{z^{(1)}_{b_{l_1+l_2}}} 
\label{eq:f_extract}\\
&\ldots 
\partial_{z^{(\nz)}_{b_{l_1+\ldots+l_{\nz-1}+1}}} \ldots  \partial_{z^{(\nz)}_{b_{l_1+\ldots+l_{\nz}}}}
 \partial_{\theta^{(1)}_{b_{|\lambda|-\lambda^t_1+1}}} \ldots \partial_{\theta^{(1)}_{b_{|\lambda|}}}
f({\bf z}) \,.
\nonumber
\end{align}
The normalizations can be explained as follows.
When extracting the components $f^{a_1\ldots a_{|\lambda|}}$ 
 from the polynomial $\bar f({\boldsymbol \theta})$ all that happens is the antisymmetrization
of a tensor which is already in the antisymmetric basis. For each set of antisymmetric indices 
every generated term is the same and the normalization factor only has
to cancel the number of terms.
Going from  $f({\bf z})$ to $f^{a_1\ldots a_{|\lambda|}}$ involves
 a Young projection of a tensor that is already Young symmetrized.
Therefore the normalization $H(\lambda)$ is that of the Young projectors,
which are given in \cite{Cvitanovic:2008zz}. It is computed from the
shape of $\lambda$ by a hook rule.
Write into each box of a Young diagram the number of boxes to its right and below,
including the box itself. The product of all numbers is $H(\lambda)$. For example,
\beq
H\left(\ \bt{youngt_211}\ \right) = H\left(\ \bt{youngt_211_hook}\ \right) = 6 \cdot 4 \cdot 3 \cdot 2 \,.
\eeq

As far as we are aware an explicit general formula for the projector
$\pi_{\lambda}^{a_1\ldots a_{|\lambda|},b_1\ldots b_{|\lambda|}}$
is only known for symmetric tensors \cite{Dobrev:1975ru}.
For the simplest mixed-symmetry tensor $\bts{young_11}$ the projector is \cite{Cvitanovic:2008zz}
\beq
\pi_{\bts{young_11}}^{a_1 a_{2} a_3,b_1 b_2 b_3} =
\frac{4}{3} \  {\mathord{\vcenter{\hbox{\scalebox{0.3}{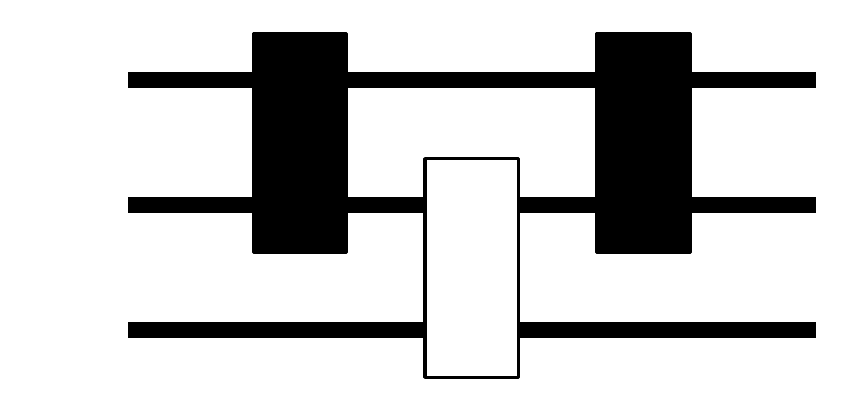}}}}} \quad
- \frac{2}{d-1} \ {\mathord{\vcenter{\hbox{\scalebox{0.3}{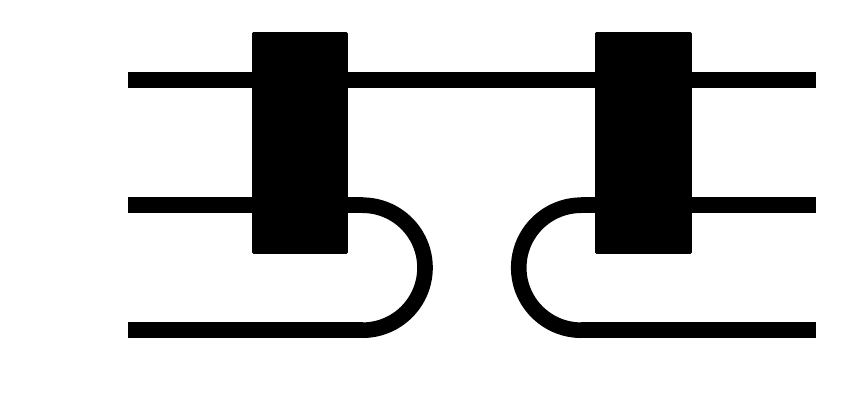}}}}} \quad.
\label{eq:irrep11_projector}
\eeq
Let $f^{a_1 a_2 a_3}$ and $g^{b_1 b_2 b_3}$ be two tensors in the irrep $\bts{young_11}$
and 
\bea
f({\bf z})=
f(z,\theta) &=
\left(z \cdot \partial_\theta\right) \theta_{a_1}\theta_{a_2}z_{a_3} f^{a_1 a_2 a_3}
= \left( \theta_{a_1} z_{a_2} z_{a_3} - \theta_{a_2} z_{a_1} z_{a_3} \right) f^{a_1 a_2 a_3} \big|_{z^2 =z\cdot \theta =\,0} \,, \\
g({\bf z})=g(z,\theta) &= 
\left(z \cdot \partial_\theta\right) \theta_{a_1}\theta_{a_2}z_{a_3} g^{a_1 a_2 a_3}
=\left( \theta_{a_1} z_{a_2} z_{a_3} - \theta_{a_2} z_{a_1} z_{a_3} \right) g^{a_1 a_2 a_3} \big|_{z^2 =z\cdot \theta =\,0} \,,
\eea{eq:poly_hook}
their encoding polynomials. We would like to know how to  contract these tensors using directly the polynomials.
The antisymmetrization in the
projector \eqref{eq:irrep11_projector} is already done in the construction of the polynomials,
only the symmetrization and subtraction of the trace is left to do. This can be done by introducing a
differential operator $D^a_z$ that satisfies
\beq
D^{a_1}_z D^{a_2}_z z^{b_1} z^{b_2} = \frac{1}{4} \left( \frac{2}{3}
\left(\delta^{a_1 b_1} \delta^{a_2 b_2} + \delta^{a_1 b_2} \delta^{a_2 b_1}\right)
- \frac{2}{d-1}\, \delta^{a_1 a_2} \delta^{b_1 b_2} \right),
\eeq
where the factor $\frac{1}{4}$ normalizes the antisymmetrizations. $D^a_z$
can be found to be
\beq
D^a_z = \frac{1}{\sqrt{6}} \left( \frac{\partial}{\partial z_a} - \frac{3}{2(d-1)} \,z^a \frac{\partial^2}{\partial z \cdot \partial z}\right).
\label{eq:todorov_op_young11_phys_space}
\eeq
The contraction of the two traceless tensors can then be expressed in terms of the encoding polynomials as
\beq
f^{a_1 a_2 a_3} g_{a_1 a_2 a_3}
= f(D_z,\partial_\theta) \,g(z,\theta) \,.
\label{eq:y11_contraction}
\eeq
This is entirely analogous to the situation of symmetric traceless tensors, but now the explicit form of the projector and corresponding 
differential operator acting on the polarization vectors is not known in general.
We will assume that there exists for every irrep $\lambda$ a set of differential operators
\beq
{\bf D}_{\bf z} = \left( D^{(1)}_{z^{(1)}} , \ldots, D^{(\nz)}_{z^{(\nz)}}, D^{(\nz+1)}_{\theta^{(1)}}  \right) ,
\eeq
that reproduces the projector in this way. We have no proof that every
projector can be expressed like this. If nothing else
it is a notation that allows us to write any contraction as
\beq
f^{a_1 \ldots a_{|\lambda|}} g_{a_1 \ldots a_{|\lambda|}}
= f({\bf D}_{\bf z} ) g({\bf z}) \,.
\eeq
We postpone a more general treatment of the projectors to traceless mixed-symmetry tensors to a subsequent paper.

\subsection{Tensors in embedding space}

To work out the constrains conformal symmetry imposes on correlation functions of tensor operators, it is convenient to use the embedding formalism. The idea, which 
dates back at least to Dirac \cite{Dirac:1936fq}, is to lift the problem to the embedding space ${\mathbb M}^{d+2}$ where the conformal group $SO(d+1,1)$ acts linearly as standard
Lorentz transformations in $(d+2)$-dimensional Minkowski space. Let  $P\in {\mathbb M}^{d+2}$ be a point in this embedding space. Points in physical space
are identified with light-rays, i.e.\ with null vectors in ${\mathbb M}^{d+2}$ up to rescalings, 
\beq
P^2=0\,,\quad\quad P\sim\alpha P \quad(\alpha>0)\,.
\eeq
Then, a specific choice of conformal frame corresponds to a specific  section of the light cone. In particular, for a CFT on $d$-dimensional Euclidean space ${\mathbb R}^{d}$,
we consider the Poincar\'e section of the light-cone 
\beq
P^A=\big(P^+,P^-,P^a\big) = \big(1,x^2,x^a\big)\,,
\label{eq:Poincare}
\eeq
where we are using light-cone coordinates with metric 
\beq
P_1\cdot P_2= \eta_{AB} P_1^AP_2^B=-\frac{1}{2}\big(P_1^+P_2^- + P_1^-P_2^+\big)+\delta_{ab} P_1^a P_2^b\,.
\eeq
For example, it is simple to see that the Euclidean distance between two points in ${\mathbb R}^{d}$ is written in the embedding space as  $-2P_1\cdot P_2 = (x_1-x_2)^2$.
It will later be abbreviated by $P_{ij} \equiv -2 P_i \cdot P_j$. 
In general, $SO(d+1,1)$ Lorentz transformations map the light-cone into itself and, by the identification (\ref{eq:Poincare}), define the action of the conformal group in physical
space. A more thorough discussion of the embedding formalism can be seen in \cite{Cornalba:2009ax,Costa:2011mg}, whose notation we follow here.  

Let us now consider a mixed-symmetry tensor  primary field of dimension $\Delta$. This field will have components 
$f^{a_1\ldots a_{|\lambda|}}(x)$ with symmetries given by the Young diagram $\lambda$. 
We wish to express it in terms of a field  on the embedding space. This new  
tensor field will have components $F^{A_1\ldots A_{|\lambda|}}(P)$ with the same symmetries as the physical tensor,
it should be defined on the light cone $P^2=0$ and it should be homogeneous of degree $-\Delta$,
\beq
F_{A_1\ldots A_{|\lambda|}} (\alpha P) = \alpha^{-\Delta} F_{A_1\ldots A_{|\lambda|}} ( P)\,, \quad \alpha>0\,.
\eeq
It should also obey the transversality condition
\beq
P^{A_i} F_{A_1\ldots A_{i} \ldots A_{|\lambda|}} =0\,.
\eeq
Components of the physical tensor are then obtained by projecting into physical space by
\beq
f_{a_1\ldots a_{|\lambda|}} = \frac{\partial P^{A_1}}{\partial x^{a_1}} \ldots \frac{\partial P^{A_{|\lambda|}}}{\partial x^{a_{|\lambda|}}} F_{A_1\ldots A_{|\lambda|}} \,.
\label{eq:physical_embedding_proj}
\eeq

Next we wish to encode the  tensor in the embedding space $F^{A_1\ldots A_{|\lambda|}}(P)$ by a polynomial. The discussion is entirely  analogous to that of the previous section, 
only that now the tensor will be a polynomial $F(P,{\bf Z})$ in the embedding space polarization vectors
\beq
{\bf Z} \equiv \left(Z^{(1)},Z^{(2)},\ldots,Z^{(\nz)},\Theta^{(1)}\right).
\eeq
Explicitly, the polynomial $F(P,{\bf Z})$ is given by
\bea
F(P,{\bf Z})
\equiv{}
\prod\limits_{p=1}^{\nz} \prod\limits_{q=1}^{\min ( l_p,\ntheta)}
\left( Z^{(p)} \cdot \partial_{\Theta^{(q)}} \right)
\bar F(P,{\boldsymbol  \Theta})  \,,
\eea{eq:def_poly_asym_embedding}
where
\bea
\bar F(P,{\boldsymbol  \Theta}) 
\equiv{}&
\Theta^{(1)}_{A_1} \ldots \Theta^{(1)}_{A_{h_1}} \Theta^{(2)}_{A_{h_1+1}} \ldots \Theta^{(2)}_{A_{h_1+h_2}} \\
&\ldots
\Theta^{(\ntheta)}_{A_{h_1+\ldots+h_{\ntheta-1}+1}} \ldots  \Theta^{(\ntheta)}_{A_{h_1+\ldots+h_{\ntheta}}} 
Z^{(1)}_{A_{|\lambda|-\lambda_1+1}} \ldots Z^{(1)}_{A_{|\lambda|}}
F^{A_1\ldots A_{|\lambda|}}(P) \,,
\eea{eq:Fbar_def}
with
\beq
{\boldsymbol  \Theta} \equiv \left( \Theta^{(1)},\Theta^{(2)}.\ldots,\Theta^{(\ntheta)},Z^{(1)} \right).
\label{eq:theta_z_vec_def_embedding}
\eeq
For traceless transverse tensors one can, without loss of information, drop scalar products of any two polarizations or of one polarization
and the corresponding embedding space coordinate, i.e.\
\bea
F^{A_1\ldots A_{|\lambda|}}(P) \text{ traceless \& transverse } &\leftrightarrow
\left.\bar F(P,{\boldsymbol  \Theta}) \right|_{
\substack{\Theta^{(p)}\cdot\Theta^{(q)} = \Theta^{(p)}\cdot Z^{(1)} = {Z^{(1)}}^2 = 0\\ \Theta^{(p)}\cdot P = Z^{(1)}\cdot P= 0}} \,,\\
&\leftrightarrow
\left.F(P,{\bf Z}) \right|_{
\substack{Z^{(p)}\cdot Z^{(q)} = Z^{(p)}\cdot \Theta^{(1)} = 0\\ Z^{(p)}\cdot P = \Theta^{(1)}\cdot P= 0}} \,.
\eea{eq:traceless_transverse}
This means that transverse polynomials satisfy the transversality condition
\beq
F(P, {\bf Z} + {\bf c} P) = F(P, {\bf Z})\,,
\eeq
for any set ${\bf c}=(c_1, \ldots, c_{\nz}, \gamma)$ of $\nz$ commuting numbers $c_i$ and one anti-commuting 
number $\gamma$.

It is also possible to relate the polynomial $f(x,{\bf z})$ to the embedding polynomial $F(P,{\bf Z})$, as well as 
$\bar{f}(x,{\boldsymbol  \theta})$  to $\bar{F}(P,{\boldsymbol  \Theta})$. The procedure is entirely analogous to that described in \cite{Costa:2011mg}:
in the case of the Poincar\'e patch where $P_x=(1,x^2,x)$, each embedding polarization can be written as
\beq
Z^{(p)}_{z,x}=\left(0,2x\cdot z^{(p)},z^{(p)}\right)
\quad {\rm and} \quad 
\Theta^{(p)}_{\theta,x}=\left(0,2x\cdot \theta^{(p)},\theta^{(p)}\right),
\label{polarization_maps}
\eeq
so that the relation between the polynomials is simply
\beq
f(x,{\bf z})= F\big(P_x, {\bf Z}_{z,x}\big)
\quad {\rm and} \quad 
\bar{f}(x,{\boldsymbol  \theta})= \bar{F}\big(P_x, {\boldsymbol  \Theta}_{\theta,x}\big)\,.
\label{polynomial_maps}
\eeq

The projector $\pi_{\lambda}^{a_1\ldots a_{|\lambda|},b_1\ldots b_{|\lambda|}}$ to the irrep $\lambda$ 
lifts to the projector $\Pi_{\lambda}^{A_1\ldots A_{|\lambda|},B_1\ldots B_{|\lambda|}}$ in embedding
space. The only case we need here is when it is inserted between two transverse tensors,
since we will always work with polynomials that are transverse.\footnote{
The general form of $\Pi_\lambda$ can be obtained analogously as it was done for
symmetric tensors in \cite{Costa:2011mg}.
}
In this case $\Pi_{\lambda}^{A_1\ldots A_{|\lambda|},B_1\ldots B_{|\lambda|}}$
is obtained from $\pi_{\lambda}^{a_1\ldots a_{|\lambda|},b_1\ldots b_{|\lambda|}}$
by replacing all Kronecker deltas $\delta^{a_i b_j}, \delta^{a_i a_j}$ and $\delta^{b_i b_j}$
by embedding space metrics $\eta^{A_i B_j}, \eta^{A_i A_j}$ and $\eta^{B_i B_j}$.
This implies that the operators ${\bf D}_{\bf z}$
can also be carried over to embedding space by
replacing ${\bf z}$ by ${\bf Z}$, when they are used between
two transverse polynomials.
The contraction of two traceless transverse tensors 
$F^{A_1 A_2 A_3}$ and $G^{B_1 B_2 B_3}$ in the irrep $\bts{young_11}$
is, as in the example \eqref{eq:y11_contraction}, given by
\beq
F^{A_1 A_2 A_3} G_{A_1 A_2 A_3}
= F(D_Z,\partial_\Theta) G(Z,\Theta)\,,
\label{eq:y11_contraction_embedding}
\eeq
with
\beq
D^A_Z = \frac{1}{\sqrt{6}} \left( \frac{\partial}{\partial Z_A} - \frac{3}{2(d-1)} \,Z^A \frac{\partial^2}{\partial Z \cdot \partial Z}\right).
\label{eq:todorov_op_young11}
\eeq
In general, contractions will be written as
\beq
F^{A_1 \ldots A_{|\lambda|}} G_{A_1 \ldots A_{|\lambda|}}
= F({\bf D}_{\bf Z}) \,G({\bf Z}) \,.
\label{eq:contraction_embedding}
\eeq

\section{Correlation functions}
\label{sec:correlation_functions}

In this section we address the main kinematic problem that is to be solved when thinking about correlation functions of arbitrary tensor irreps: to count, and to construct,  all independent tensor structures. 

\subsection{Tensor-product coefficients}

One part of the problem is finding
all the possible ways a given set of mixed-symmetry tensors can be contracted.
A more mathematical way to pose this question is to ask for the multiplicity of the scalar representation in the tensor product
of the tensors in question. Fortunately, this problem is already solved. Here we shall review the relevant results for our purposes;
for a comprehensive introduction to the general properties of tensor-product coefficients see \cite{DiFrancesco:1997nk}.

Let $G$ be $SU(n)$, $SO(n)$ or $Sp(n)$ and $\lambda, \mu, \nu$ irreducible $G$-modules which are enumerated by Young diagrams. These are the vector spaces of
tensors with the index symmetries described in Section \ref{sec:index_free_notation}. They will often be called representations instead of modules
in the following. $\lambda^{*}$ denotes the vector space dual to $\lambda$, i.e.\ if $\lambda$ contains tensors with lower indices,
$\lambda^{*}$ contains tensors with upper indices. Upper and lower indices can be contracted and the result will then transform under $G$
as indicated by the remaining indices.

Let $\mathcal{N}_{\lambda \mu}^{\ \;\; \nu}$ be the tensor-product coefficients of $G$.
They count the multiplicity with which the irrep $\nu$ appears in the tensor product of $\lambda$ and $\mu$
\beq
\lambda \otimes \mu = \bigoplus\limits_\nu \mathcal{N}_{\lambda \mu}^{\ \;\; \nu} \,\nu \,,
\label{eq:tensor_product}
\eeq
and satisfy
\beq
\mathcal{N}_{\lambda \bullet}^{\ \;\; \nu} = \delta_\lambda^\nu \,,\quad\quad
 \mathcal{N}_{\lambda \lambda^*}^{\ \ \  \bullet} = 1\,, \quad\quad
\mathcal{N}_{\lambda \mu}^{\ \;\; \nu} = \mathcal{N}_{\lambda \nu^*}^{\ \ \  \mu^*} \,,
\label{eq:tensor_product_singlet}
\eeq
where $\bullet$ denotes the scalar representation.
Let us also denote by $\mathcal{N}_{\lambda \mu \nu}$  the multiplicity of the scalar representation in the triple product
\beq
\lambda \otimes \mu \otimes \nu = \mathcal{N}_{\lambda \mu \nu} \bullet \,\oplus \text{ other irreps} \,.
\eeq
This notation has the advantage of being symmetric in its three labels and contains the same information due to
\beq
\mathcal{N}_{\lambda \mu}^{\ \;\; \nu} = \mathcal{N}_{\lambda \mu \nu^*} .
\eeq
The multiplicity of a given representation $\mu$ in products of more than two tensors will be denoted by
$\mathcal{N}_{\lambda_1 \ldots \lambda_n}^{\qquad  \mu}$
\beq
\lambda_1 \otimes \ldots \otimes \lambda_n = \bigoplus\limits_{\mu} \mathcal{N}_{\lambda_1 \ldots \lambda_n}^{\qquad \mu}\, \mu \, ,
\eeq
and can be calculated by recursively using \eqref{eq:tensor_product}
\beq
\mathcal{N}_{\lambda_1 \ldots \lambda_n}^{\qquad \mu} =
\sum\limits_{\nu_3,\ldots,\nu_n}
\mathcal{N}_{\lambda_1 \lambda_2}^{\quad \  \nu_3}
\prod\limits_{i=3}^{n-1}
\left(
\mathcal{N}_{\nu_i \lambda_i}^{\quad \  \nu_{i+1}}
\right)
\mathcal{N}_{\nu_n \lambda_n}^{\quad \  \mu} \, .
\label{eq:higher_tp_coeff}
\eeq
This also computes the multiplicity of the scalar representation in the product $\lambda_1 \otimes \ldots \otimes \lambda_n \otimes\mu^*$.

\subsubsection{Unitary groups}

When specializing to $G = SU(n)$ the tensor-product coefficients are the famous Littlewood-Richardson coefficients $c_{\lambda \mu}^{\ \;\; \nu}$,
\beq
\mathcal{N}_{\lambda \mu}^{\ \;\; \nu} = c_{\lambda \mu}^{\quad \nu} \quad \text{for } G = SU(n) \,.
\eeq
The only allowed contraction in this group is between upper and lower indices, so the number of indices adds up when
the tensor product between two tensors with lower indices is formed
\beq
c_{\lambda \mu}^{\quad \nu} = 0 \quad \text{for } |\lambda| + |\mu| \neq |\nu| \,.
\eeq
This implies that the product of three tensors can only contain the scalar representation if one of them is in a
dual representation relative to the other two. This can be illustrated by the following schematic contraction of tensor indices
\beq
{\mathord{\vcenter{\hbox{\scalebox{0.3}{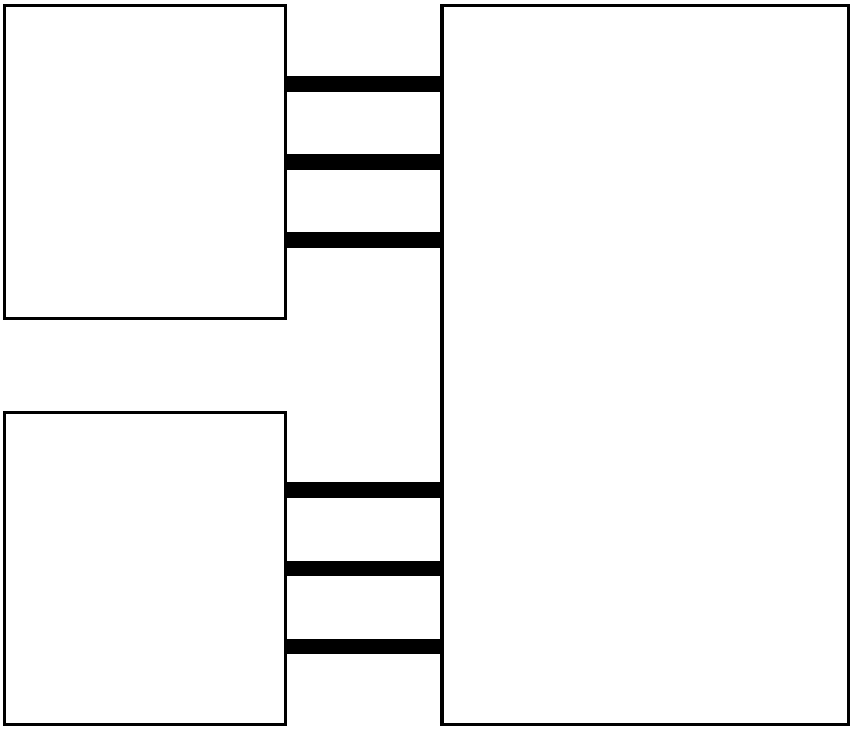}}}}} \ \propto \  c_{\lambda \mu}^{\quad \nu} \bullet \,.
\eeq
The coefficients $c_{\lambda \mu}^{\quad \nu}$ can be calculated using the Littlewood-Richardson
rule \cite{LR}.\footnote{The algorithm has been implemented for instance in
Anders Skovsted Buch's lrcalc program, which is available at
\url{http://www.math.rutgers.edu/~asbuch/lrcalc/}.}

For simple examples one can often find the possible contractions for a given tensor product
quickly using birdtracks. For example, one can easily convince oneself that the only
two inequivalent ways to contract $\lambda = \mu = \bts{young_11}$ and $\nu^* = \bts{young_111}^*$
are
\beq
{\mathord{\vcenter{\hbox{\scalebox{0.3}{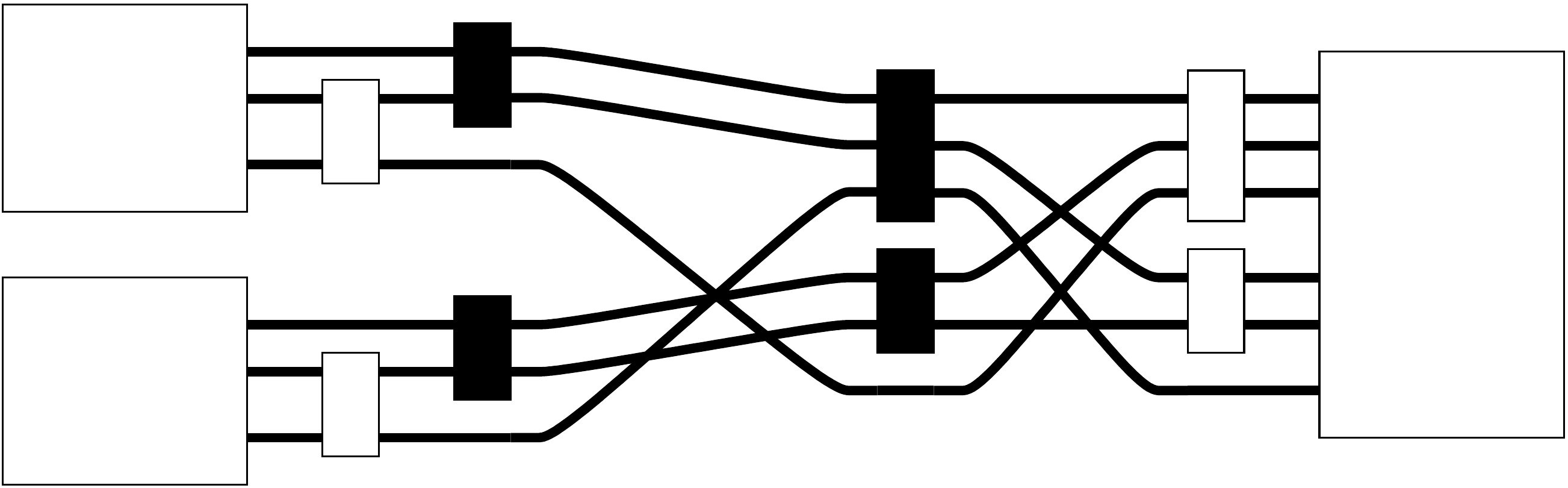}}}}} \;,
\eeq
and
\beq
{\mathord{\vcenter{\hbox{\scalebox{0.3}{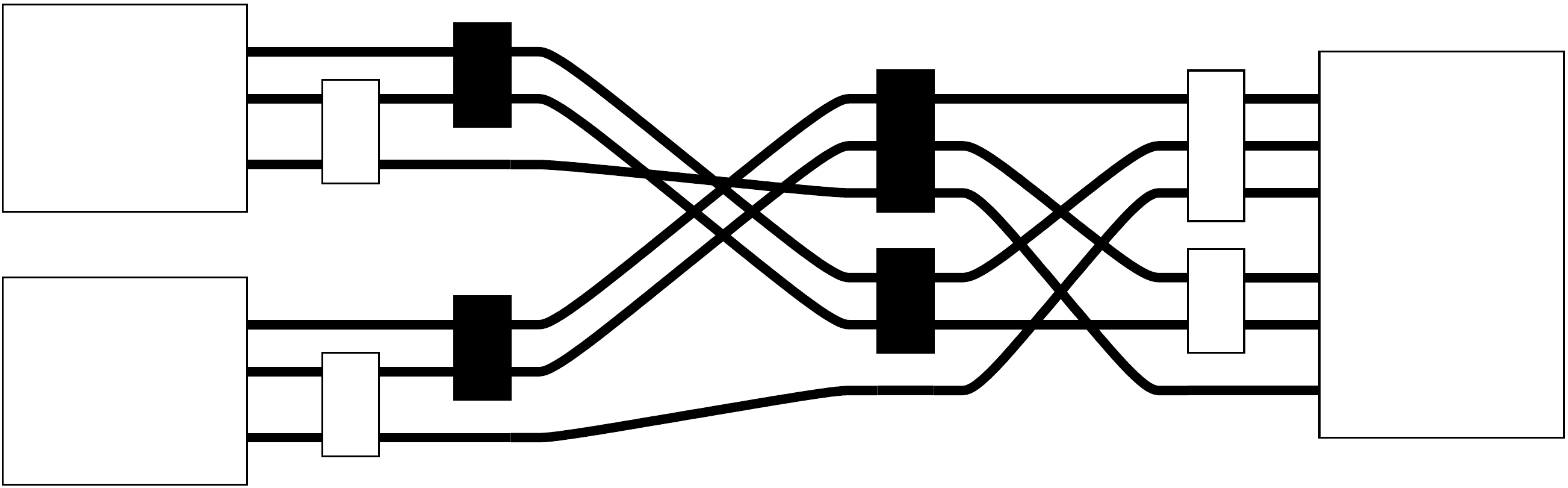}}}}} \;.
\eeq
The Littlewood-Richardson coefficient is thus $c_{\lambda \mu}^{\quad \nu} = 2$.

\subsubsection{Orthogonal and symplectic groups}
\label{sec:SO_LR_coeffs}

Following the reasoning of \cite{Cvitanovic:2008zz}, the orthogonal and symplectic groups can be obtained
from the unitary groups by taking into account the fact that these groups have by definition additional group invariants.
For $SO(d)$ this is a symmetric quadratic form $g_{ab}$ and its inverse $g^{ab}$, while for $Sp(d)$ the invariant is skew symmetric
$f_{ab}=-f_{ba}$. In both cases these invariants can be used to raise and lower indices, which implies that the distinction between
the two becomes unnecessary, the representations are self-dual $\lambda^* = \lambda$. Any two indices can be contracted and this leads to different tensor-product coefficients
\beq
\mathcal{N}_{\lambda \mu}^{\ \;\; \nu} =
\mathcal{N}_{\lambda \mu \nu} = b_{\lambda \mu \nu}\quad \text{for } G \in \big\{ SO(2n), SO(2n+1), Sp(2n)\big\} \,.
\label{eq:SO_SP_LR_coeffs}
\eeq
Because of the self-duality of the representations the position of the indices of the tensor-product coefficients becomes meaningless,
so these coefficients are always written with only lower indices.
It is not hard to convince oneself that the counting of tensor structures here can be broken down to the counting that was relevant in the $SU(n)$
case where the restriction $|\lambda| + |\mu| = |\nu|$ applied. The following figure shows how three sets of indices can be
contracted with each other, by first dividing each set of indices into two,
\beq
\sum_{\rho, \sigma, \gamma} \quad {\mathord{\vcenter{\hbox{\scalebox{0.3}{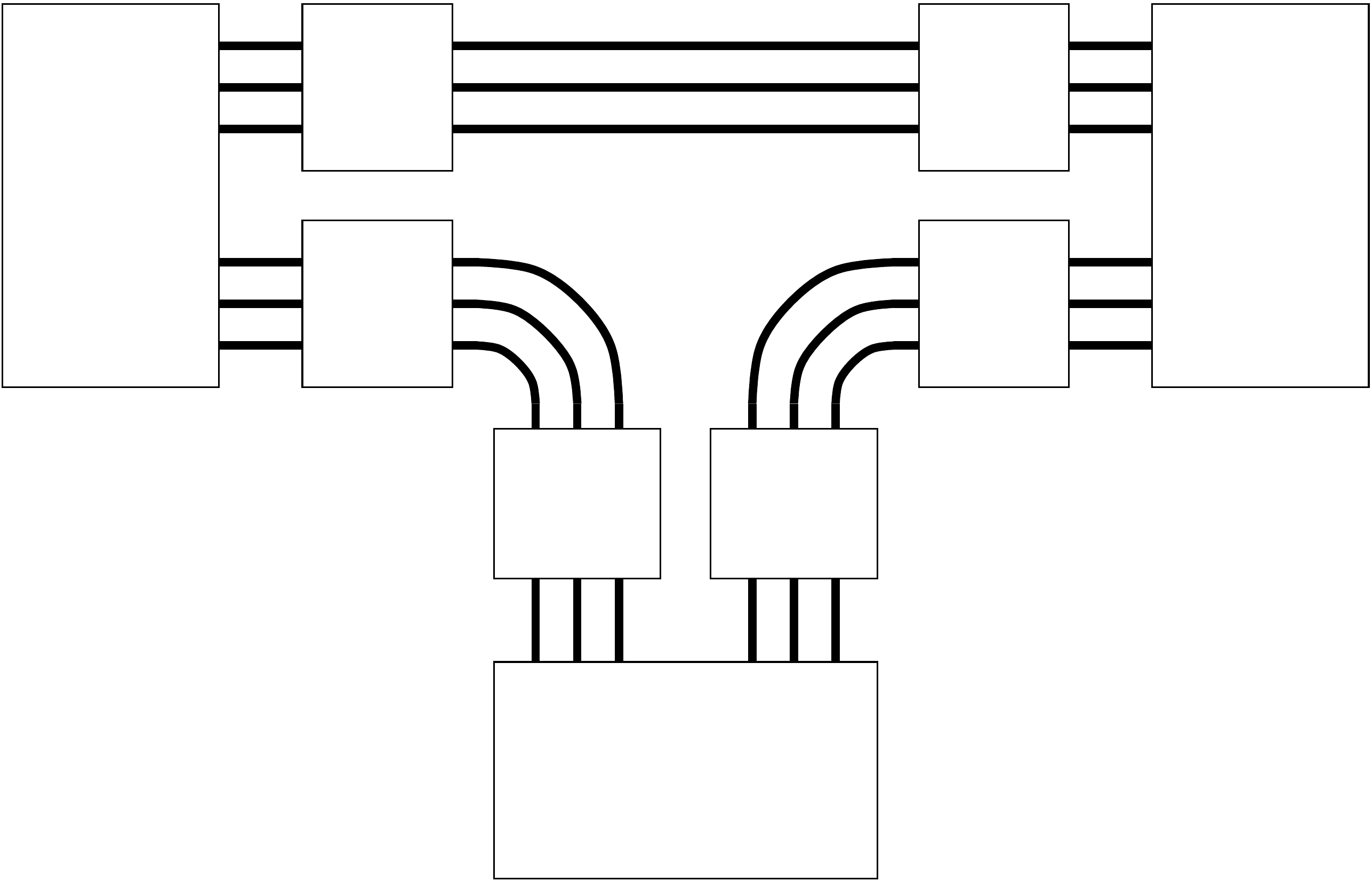}}}}} \ \propto \  b_{\lambda \mu \nu}\, \bullet \,,
\eeq
where $\mathbf{P}_\rho$ is a projector to the irrep $\rho$, and so on.
The number of tensor structures obtained in such a way is
\beq
b_{\lambda \mu \nu}
=
\sum_{\rho, \sigma, \gamma} c^{\quad \lambda}_{ \sigma \rho} \,c^{\quad \mu}_{\rho \gamma} \,c^{\quad \nu}_{\gamma \sigma} .
\label{eq:newell-littlewood}
\eeq
This formula is known as the Newell-Littlewood formula \cite{MR0095209, MR991410} and
holds if the sum of the heights of two of the three irreps $\lambda, \mu$ and $\nu$ does not exceed $n$, i.e.\ 
for
\beq
h_1^\lambda + h_1^\mu + h_1^\nu - \max \left(h_1^\lambda, h_1^\mu, h_1^\nu\right) \leq n = \Big\lfloor \frac{d}{2} \Big\rfloor \,.
\label{eq:no_modification_condition}
\eeq
Otherwise even the tensor product of the two irreps with the smallest $h_1$ contains Young diagrams that
violate \eqref{eq:irrep_height_restriction} and hence
do not correspond to irreps of $SO(d)$ or $Sp(d)$.
In this case \eqref{eq:newell-littlewood} can be used anyway by transforming these Young diagrams into diagrams that
correspond to irreps using modification rules \cite{MR885807}
and taking the additional contributions that arise in this way into account.
Then also the statement \eqref{eq:SO_SP_LR_coeffs} that the tensor-product coefficients
are the same for $SO(2n)$, $SO(2n+1)$ and $Sp(2n)$ does not hold true anymore.
For simplicity, we will assume \eqref{eq:no_modification_condition} to be satisfied throughout the paper.
Note that this implies that explicit examples in this paper hold only for $d$ sufficiently large.

The coefficients descibing the decomposition of the tensor product of more than two irreps are
given by \eqref{eq:higher_tp_coeff}
\beq
b_{\lambda_1 \ldots \lambda_n} =
\sum\limits_{\nu_3,\ldots,\nu_n}
b_{\lambda_1 \lambda_2 \nu_3}
\prod\limits_{i=3}^{n-2}
\left(
b_{\nu_i \lambda_i}^{\quad \  \nu_{i+1}}
\right)
b_{\nu_{n-1} \lambda_{n-1} \lambda_n} \, .
\label{eq:higher_tp_coeff_SO}
\eeq
For $SO(d)$ or $Sp(d)$ the same coefficients also count the multiplicity of the scalar representation in the tensor product
$\lambda_1 \otimes \lambda_2 \otimes \ldots \otimes \lambda_n$.

A notation that will be used below is the restriction of a tensor product to irreps that have the same number
of indices as both irreps in the product. 
This operation will be denoted with square brackets and amounts to using the $SU(n)$ Littlewood-Richardson
coefficients as tensor-product coefficients,
\beq
[ \lambda \otimes \mu ] \equiv \bigoplus\limits_\nu b_{\lambda \mu \nu} \, \nu \Big|_{|\nu| = |\lambda| + |\mu|} 
=\bigoplus\limits_\nu c_{\lambda \mu}^{\quad \nu} \, \nu \,.
\label{eq:square_tensor_product}
\eeq
The second equality can be found for instance in \cite{MR885807}.
To wrap up this section consider the following example 
\beq
[\lambda \otimes \mu \otimes \nu] \otimes \rho \otimes \sigma
=  \left(\sum\limits_{\gamma, \kappa} c_{\lambda \mu}^{\quad \gamma} \,c_{\gamma \nu}^{\quad \kappa} \,b_{\kappa \rho \sigma} \,\bullet\right)
 \oplus \text{ other irreps}\,.
\eeq

\subsection{Two-point functions}

Unitary irreducible representations of the conformal group $SO(d+1,1)$ will be labeled by $\chi \equiv [\lambda,\Delta]$,
where $\Delta$ is the conformal dimension and $\lambda$  an irreducible representation of $SO(d)$. 
The two-point function of the primary corresponding to $\chi$ is, up to a normalization constant, a tensor depending on two points in the embedding space with components
\beq
G^{A_1\ldots A_{|\lambda|}B_1\ldots B_{|\lambda|}} \big(P_1,P_2\big) \,.
\eeq
It is encoded, as described above, by a polynomial
\bea
G_{\chi} \big(P_1,P_2;\bZ_1, \bZ_2\big)
=
\prod\limits_{p=1}^{\nz} \prod\limits_{q=1}^{\min ( l_p,\ntheta)}
\left( Z^{(p)}_1 \cdot \partial_{\Theta^{(q)}_1} \right) \left( Z^{(p)}_2 \cdot \partial_{\Theta^{(q)}_2} \right)
\bar G_{\chi} (P_1,P_2; \bTheta_1, \bTheta_2) \,,
\eea{eq:def_two-point}
where
\bea
&\bar G_{\chi} \big(P_1,P_2; \bTheta_1, \bTheta_2\big)=
\\
={}&
\Theta^{(1)}_{1A_1} \ldots \Theta^{(\ntheta)}_{1A_{|\lambda|-\lambda_1}}  \ldots Z^{(1)}_{1A_{|\lambda|}}
 \Theta^{(1)}_{2B_1} \ldots \Theta^{(\ntheta)}_{2B_{|\lambda|-\lambda_1}}  \ldots Z^{(1)}_{2B_{|\lambda|}}
G^{A_1\ldots A_{|\lambda|}B_1\ldots B_{|\lambda|}} \big(P_1,P_2\big) \,.
\eea{eq:def_F}
To construct the two-point function one has to find $\bar G_{\chi}$, which is subject to the following
conditions. Firstly, it is homogeneous of degree $-\Delta$ in the embedding space coordinates
\beq
\bar G_{\chi} \big(\{\alpha_i P_i;\bTheta_i\}\big)
= (\alpha_1 \alpha_2)^{-\Delta} \bar G_{\chi} \big(\{P_i;\bTheta_i\}\big)\,,
\label{eq:2point-scale-P}
\eeq
for $\alpha_i$ arbitrary positive constants. 
Secondly, it is a polynomial in the polarizations with degrees given by the shape of the Young diagram $\lambda$,
\beg
\bar G_{\chi} \big(\{P_i; \bbeta_i \bTheta_i)\}\big)
= \left(\beta_1^{(1)} \beta_2^{(1)}\right)^{h_1} \ldots \left(\beta_1^{(\ntheta)} \beta_2^{(\ntheta)}\right)^{h_{\ntheta}} \left(\b_1^{(Z)} \b_2^{(Z)}\right)^{\lambda_1} 
\bar G_{\chi} \big(\{P_i;\bTheta_i\}\big) \,,
\eeg{eq:2point-scale-Z}
where we defined
\beq
\bbeta_{i} \bTheta_i = \left(\beta_i^{(1)}\Theta_i^{(1)},\ldots,\beta_i^{(\ntheta)}\Theta_i^{(\ntheta)},\b_i^{(Z)}Z_i^{(1)}\right),
\eeq
for arbitrary (commuting) constants $\beta_i^{(p)}$.

Finally, $\bar G_{\chi}$ has to be transverse
\beg
\bar G_{\chi} \big(\{P_i; \bTheta_i+ \bgamma_{i}P_i\}\big)
= \bar G_{\chi} \big(\{P_i;\bTheta_i\}\big) \,.
\eeg{eq:2point-trans}
where
\beq
\bgamma_i=\left(\gamma_i^{(1)}, \ldots, \gamma_i^{(\nz)}, c_i\right),
\eeq
is a set of $\nz$ anticommuting numbers and one commuting number.
This last condition has to be satisfied modulo $O\big(P^2\big)$ terms. An identically transverse function $\bar G_{\chi}$ can be obtained by dropping
terms proportional to 
$\Theta^{(p)}\cdot\Theta^{(q)}$ and $\Theta^{(p)}\cdot P$,  where $p=1, \ldots, \ntheta, Z$.
Notice that we are using the notation $\Theta^{(Z)}=Z^{(1)}$ to make equations more compact.
We are left to constructing $\bar G_{\chi}$  from the tensors 
\beq
C^{(p)}_{iAB} =
\Theta^{(p)}_{iA} P_{iB} - \Theta^{(p)}_{iB} P_{iA}=
\begin{cases}\Theta^{(p)}_{iA} P_{iB} - \Theta^{(p)}_{iB} P_{iA} \,, 
\quad &p = 1,\ldots, \ntheta \,, \\
 Z^{(1)}_{iA} P_{iB} - Z^{(1)}_{iB} P_{iA}\,, 
 \quad &p = Z \,,
\end{cases}
\eeq
with $i=1,2$.
Contracting two such tensors with the same index $i$ leads to terms of the type that do not appear in transverse functions,
so the only possible terms are traces of a string of $C$'s with alternating $i$'s, i.e.\ of the form
\beq
\Tr \left(C^{(p)}_1 \cdot C^{(q)}_2 \ldots C^{(r)}_1 \cdot C^{(s)}_2\right) ,
\eeq
the shortest one being
\beq
H_{ij}^{(p,q)} \equiv \Tr \left(C_i^{(p)} \cdot C_j^{(q)}\right) 
= 2 \bigg( \Big(P_j \cdot \Theta_i^{(p)}\Big) \Big(P_i \cdot \Theta_j^{(q)}\Big) - \Big(\Theta_i^{(p)} \cdot \Theta_j^{(q)}\Big) \Big(P_i \cdot P_j\Big)\bigg) .
\label{eq:Hij_def}
\eeq
Recall that both $p$  and/or $q$ can also take the value $Z$, for which case they describe the commuting polarization $Z^{(1)}$.

Traces of more than two alternating $C_1$'s and $C_2$'s can always be expressed in terms of $H_{12}^{(p,q)}$. This can be seen by considering
\beg
\left(C^{(p)}_{1} \cdot C^{(q)}_{2}\cdot C^{(r)}_{1} \cdot C^{(s)}_{2}\right)_{AB}
= \frac{1}{2} \left( C^{(p)}_{1AC} H_{21}^{(q,r)} C^{(s)}_{2CB} + P_{1A} P_{2B} R\right) ,
\eeg{eq:CCCC}
where
\begin{align}
R={}& \Theta^{(p)}_{1A} H_{21}^{(q,r)} \Theta^{(s)}_{2A} + C^{(p)}_{1AB}  \left(\Theta^{(q)}_{2}\cdot \Theta^{(r)}_{1}\right) C^{(s)}_{2BA}
-\left(\Theta^{(p)}_{1}\cdot \Theta^{(q)}_{2}\right) H_{12}^{(r,s)}
- H_{12}^{(p,q)}\left(\Theta^{(r)}_{1}\cdot \Theta^{(s)}_{2}\right)
\nonumber\\
&+2 \big(P_1 \cdot P_2\big) \left[\Theta^{(p)}_{1A} \left(\Theta^{(q)}_{2}\cdot \Theta^{(r)}_{1}\right) \Theta^{(s)}_{2A} 
-\left(\Theta^{(p)}_{1}\cdot \Theta^{(q)}_{2}\right) \left(\Theta^{(r)}_{1}\cdot \Theta^{(s)}_{2}\right) \right] ,
\label{eq:R}
\end{align}
which satisfies
\beq
\big(P_1 \cdot P_2\big) R = \frac{1}{2} \left(H_{12}^{(p,q)} H_{12}^{(r,s)}
-C^{(p)}_{1AB} H_{21}^{(q,r)} C^{(s)}_{2BA}  \right) .
\eeq
Using also that
\beq
\left( P_2 \cdot C^{(p)}_{1} \cdot C^{(q)}_{2} \right)_A = \frac{1}{2} \,H_{12}^{(p,q)} P_{2A}\,,
\label{eq:PCC=HP}
\eeq
one sees that multiplying \eqref{eq:CCCC} by any number of factors $C_1 \cdot C_2$
produces only more terms of the same structure that turn into products of $H_{12}$'s when the trace is closed.

Naively one could imagine that the different ways to distribute polarizations among $H_{12}$'s
lead to different tensor structures, e.g.\ for the diagram $\bts{young_02}$ one could consider
\beq
\left( H_{12}^{(1,1)} H_{12}^{(2,2)} \right)^2 , \quad H_{12}^{(1,1)} H_{12}^{(2,2)} H_{12}^{(1,2)} H_{12}^{(2,1)}  \quad\text{ and } \quad \left( H_{12}^{(1,2)} H_{12}^{(2,1)} \right)^2 .
\eeq
However, the tensor product of two copies of an irrep contains the scalar representation with multiplicity one, as written in  \eqref{eq:tensor_product_singlet},
so there can be only one tensor structure for each two-point function.
Indeed, all possible ways to distribute the polarizations among the $H_{12}$'s lead to the same result
after Young symmetrization (this can be checked explicitly by considering (\ref{eq:def_two-point})).
With the weights of coordinates and polarizations being fixed by \eqref{eq:2point-scale-P} and \eqref{eq:2point-scale-Z},
we choose a convenient set of $H_{12}$'s and find that the unique tensor structure for the two-point function is given by (\ref{eq:def_two-point}) with
\beq
\bar G_{\chi} \big(P_1,P_2; \bTheta_1, \bTheta_2\big)= \frac{1}{(P_{12})^{\Delta+|\lambda|}}\,   \prod\limits_{r=1}^{\ntheta}\left( H_{12}^{(r,r)} \right)^{h_r} \left( H_{12}^{(Z,Z)} \right)^{\lambda_1}\,.
\label{eq:result_two-point}
\eeq

\subsubsection{Example: $p$-form field}

As an example, let us write explicitly the two-point function of a $p$-form field. The Young diagram of a $p$-form field
consists of one column of $p$ boxes, therefore \ $|\lambda| = p$, $\nz = 0$ and $\ntheta = 1$.
Since there are no rows with more than one box and hence there are no indices to symmetrize, there is no need to introduce commuting polarizations. 
There is a single anti-commuting polarization vector, which we denote by $\Theta$.
The correlation function can be read off from \eqref{eq:result_two-point} to be
\beq
\bar{G}_{\chi} \big(P_1,P_2;\Theta_1,\Theta_2\big)=
\frac{\left( H_{12}^{(\Theta,\Theta)} \right)^{p}}
{(P_{12})^{\Delta+p}} =
\frac{1}{(P_{12})^\Delta}
 \left( \big(\Theta_1 \cdot \Theta_2\big) - \frac{\big(P_2 \cdot \Theta_1\big) \big(P_1 \cdot \Theta_2\big)}{P_1 \cdot P_2} \right)^p\,.
\label{eq:ex_p_form_2point}
\eeq
Then, using the maps  (\ref{polarization_maps}) and (\ref{polynomial_maps}), 
it is simple to find the polynomial $\bar{g}_\chi(x_1,x_2;\theta_1,\theta_2)$ that describes this tensor structure in physical space.

Note also that, acting with the $\Theta$ derivatives $\partial_{\Theta_1^{A_1}}\ldots \partial_{\Theta_1^{A_p}} \partial_{\Theta_2^{B_1}}\ldots \partial_{\Theta_2^{B_p}}$,
one can write explicitly the components of the tensor in the embedding space as 
\beq
G_\chi^{A_1 \ldots A_p B_1\ldots B_p} = \frac{1}{(P_{12})^\Delta}\,
\delta^{A_1}_{[C_1} \dots\delta^{A_p}_{C_p]}  \,\delta^{B_1}_{[D_1} \dots\delta^{B_p}_{D_p]}  \prod\limits_{k=1}^{p} \left(  \eta^{C_kD_k} -\frac{P_2^{C_k} P_1^{D_k}}{P_1 \cdot P_2}\right),
\eeq
whose projection to physical space gives the components
\beq
g_\chi^{a_1 \ldots a_p b_1\ldots b_p} =\frac{1}{(x_{12}^2)^{\Delta}}\,
\delta^{a_1}_{[c_1} \dots\delta^{a_p}_{c_p]}  \,\delta^{b_1}_{[d_1} \dots\delta^{b_p}_{d_p]}  \prod\limits_{k=1}^{p} \left(  \delta^{c_kd_k} - 2\, \frac{(x_{12})^{c_k}(x_{12})^{d_k}}{x_{12}^2}  \right),
\eeq
where $x_{12}=x_1-x_2$.

\subsubsection{Example: Smallest hook diagram}

As another example let us  consider the  irrep corresponding to the diagram $\bts{young_11}$. This is 
the simplest example where the Young symmetrization operator appears.
Here we have $\nz = 1$ and $\ntheta = 1$, with polarization vectors  ${\bf Z}=(Z,\Theta)$ and $\bTheta=(\Theta,Z)$.
Thus, the polynomials encoding the tensor structure for the two-point function of these operators are
\beq
\bar{G}_{\chi} \big(P_1,P_2;\bTheta_1,\bTheta_2\big) = \frac{1}{(P_{12})^{\Delta+3}}\,\left( H_{12}^{(\Theta,\Theta)} \right)^2H_{12}^{(Z,Z)} \,,
\eeq
and
\bea
G_\chi \big(P_1,P_2;{\bf Z}_1,{\bf Z}_2\big) &=  \left( Z_1 \cdot \partial_{\Theta_1} \right) \left( Z_2 \cdot \partial_{\Theta_2} \right)
\bar{G}_{\chi} \big(P_1,P_2;\bTheta_1,\bTheta_2\big)\\
&=\frac{2}{(P_{12})^{\Delta+3}}  \left(
H_{12}^{(\Theta,\Theta)}
 H_{12}^{(Z,Z)} - H_{12}^{(\Theta,Z)} H_{12}^{(Z,\Theta)} 
\right) H_{12}^{(Z,Z)} \,.
\eea{eq:ex_hook_2point}
Using the differential operator \eqref{eq:todorov_op_young11},
it is a simple exercise to derive the components of the physical tensor associated to this polynomial,
which were already derived in \cite{Alkalaev:2012rg,Alkalaev:2012ic} for all hook shaped Young diagrams.
We shall not pursue this here, and work instead with embedding
polynomials.

\subsection{Three-point functions}
\label{sec:three-point_functions}

Next we consider the tensor structures allowed in a three-point functions with each operator in the $SO(d+1,1)$ irrep labelled by
$\chi_j \equiv [\lambda_j,\Delta_j]$, for  $j=1,2,3$. Such three-point function is conveniently written as
\bea
G_{\chi_1 \chi_2 \chi_3} \big(\{P_i;\bZ_i\}\big)
=
\prod\limits_{j=1}^{3}
\prod\limits_{p=1}^{\nz^j} \prod\limits_{q=1}^{\min ( l_p^j,\ntheta^j)}
\!\!
\left( Z^{(p)}_j \cdot \partial_{\Theta^{(q)}_j} \right) 
\frac{\bar Q_{\lambda_1\lambda_2 \lambda_3} \big(\{P_i;\bTheta_i\}\big)}
{(P_{12})^{\frac{\tau_1+\tau_2-\tau_3}{2}} (P_{23})^{\frac{\tau_2+\tau_3-\tau_1}{2}} (P_{31})^{\frac{\tau_3+\tau_1-\tau_2}{2}}} ,
\eea{eq:def_three-point}
where $\tau_i = \Delta_i + |\lambda_i|$. The factor in the denominator was included to 
give $\bar Q_{\lambda_1 \lambda_2 \lambda_3}$ the same overall weight in embedding space coordinates as in polarizations, 
therefore simplifying its construction out of building blocks that have the same property.
The conditions on  $\bar Q_{\lambda_1 \lambda_2\lambda_3}$ are otherwise analogous to (\ref{eq:2point-scale-P}-\ref{eq:2point-trans}), i.e.\ 
\bea
&\bar Q_{\lambda_1\lambda_2 \lambda_3} \big(\{\alpha_i P_i;  \bbeta_{i} (\bTheta_i+ \bgamma_{i}P_i) \}\big)=
\\
={}& \bar Q_{\lambda_1 \lambda_2 \lambda_3} \big(\{P_i;\bTheta_i\}\big) \prod\limits_i \alpha_i^{|\lambda_i|} 
\left(\beta_i^{(1)}\right)^{h^i_1} \ldots \left(\beta_i^{(\ntheta^i)} \right)^{h^i_{\ntheta^i}} \left( \beta_i^{(Z)}\right)^{(\lambda_i)_1} 
\,.
\eea{eq:3point-scale}

In addition to $H_{ij}^{(p,q)}$ given in \eqref{eq:Hij_def}, there are now other building blocks that can appear in the polynomial 
$\bar Q_{\lambda_1 \lambda_2 \lambda_3}\big(\{P_i;\bTheta_i\}\big)$,
which are
\bea
V_{i,jk}^{(p)} &\equiv \frac{P_j \cdot C_i^{(p)}\cdot P_k}{P_j \cdot P_k}
=\frac{\Big(P_j \cdot \Theta^{(p)}_{i}\Big)\Big( P_{i} \cdot P_k \Big)- \Big(P_j \cdot P_{i}\Big)\Big( \Theta^{(p)}_{i} \cdot P_k\Big)}{P_j \cdot P_k} \,.
\eea{eq:def_Vijk}
Because of the property $V_{i,jk}^{(p)} = - V_{i,kj}^{(p)}$ there is only one independent $V^{(p)}$ for each operator $i$. They will be denoted
\beq
V_1^{(p)} = V_{1,23}^{(p)} \,, \qquad V_2^{(p)} = V_{2,31}^{(p)} \,, \qquad V_3^{(p)} = V_{3,12}^{(p)} \,.
\eeq
Other terms of the form $P \cdot C\cdot  \ldots \cdot C \cdot P$ can always be expressed in terms of $V_{i}^{(p)}$ and $H_{ij}^{(p,q)}$ due to
\eqref{eq:PCC=HP}. One could imagine that traces of more than two $C$'s result in independent terms, but it was proven in \cite{Costa:2011mg}
that this is not the case. This means that parity invariant three-point functions can be completely constructed out of $V_{i}^{(p)}$ and $H_{ij}^{(p,q)}$.\footnote{
Additional parity odd tensor structures can be constructed using the fully antisymmetric $\epsilon$-tensor \cite{Costa:2011mg}.
The number of such tensor structures depends on the dimension and is not considered here.
}

Let us first consider the terms in the polynomial $\bar Q_{\lambda_1\lambda_2 \lambda_3} \big(\{P_i;\bTheta_i\}\big)$ 
that are constructed only out of $H_{ij}^{(p,q)}$'s.
The number of independent structures that can arise from such terms is given by the tensor product
coefficient $b_{\lambda_1\lambda_2\lambda_3}$ of $SO(d)$, which was introduced in Section \ref{sec:SO_LR_coeffs}.
We shall denote by $\bar{W}_{\lambda_1\lambda_2\lambda_3}$ the linear combination (with arbitrary coefficients) of these 
$b_{\lambda_1\lambda_2\lambda_3}$ combinations of $H_{ij}^{(p,q)}$'s
that lead to independent tensor structures and scale as in \eqref{eq:3point-scale}.
Such a function can easily be constructed for any example by constructing terms and checking if they give rise to independent
tensor structures after the full Young symmetrization.

As an example consider one of the first combinations of irreps where the Littlewood-Richardson coefficient
is larger than one, $\lambda_1 = \lambda_2 = \bts{young_11}\,$, $\lambda_3 = \bts{young_2}\,$. The corresponding
Littlewood-Richardson coefficient is $b_{\, \bts{young_11}\,\bts{young_11}\,\bts{young_2}} = 2$.
Indeed, there are two combinations of the $H_{ij}^{(p,q)}$
that lead to different tensor structures
\bea
\left( Z_1 \cdot \partial_{\Theta_1} \right)
\left( Z_2 \cdot \partial_{\Theta_2} \right)&
\left( H_{12}^{(\Theta,\Theta)} \right)^2 H_{13}^{(Z,Z)} H_{23}^{(Z,Z)}= \\
&={} 2\left(H_{12}^{(Z,Z)} H_{12}^{(\Theta,\Theta)} - H_{12}^{(\Theta,Z)} H_{12}^{(Z,\Theta)} \right)H_{13}^{(Z,Z)} H_{23}^{(Z,Z)} ,\\
\left( Z_1 \cdot \partial_{\Theta_1} \right)
\left( Z_2 \cdot \partial_{\Theta_2} \right)&
H_{12}^{(Z,Z)} H_{12}^{(\Theta,\Theta)}  H_{13}^{(\Theta,Z)} H_{23}^{(\Theta,Z)} =\\
&={} H_{12}^{(Z,Z)} \left[\left(
H_{12}^{(\Theta,\Theta)}  H_{13}^{(Z,Z)} 
-H_{12}^{(Z,\Theta)}  H_{13}^{(\Theta,Z)}\right) H_{23}^{(Z,Z)}\right.\\
&\left.\qquad\qquad+ \left( H_{12}^{(Z,Z)}  H_{13}^{(\Theta,Z)} 
-H_{12}^{(\Theta,Z)}  H_{13}^{(Z,Z)} \right) H_{23}^{(\Theta,Z)}\right] .
\eea{eq:LR_example}
Thus, we conclude that for this example
\beq
\bar{W}_{\, \bts{young_11}\,\bts{young_11}\,\bts{young_2}}
= c_1 \left( H_{12}^{(\Theta,\Theta)} \right)^2 H_{13}^{(Z,Z)} H_{23}^{(Z,Z)}
+  c_2 \,H_{12}^{(Z,Z)} H_{12}^{(\Theta,\Theta)}  H_{13}^{(\Theta,Z)} H_{23}^{(\Theta,Z)} \,,
\eeq
with $c_1$ and $c_2$ constants.

Next we describe how to construct the general terms containing both $H_{ij}^{(p,q)}$'s and $V_{i}^{(p)}$'s.
A given term may have an arbitrary number of $V_{i}^{(p)}$'s. However, since 
for $p \in \{1, \ldots, \ntheta \}$ the $V_{i}^{(p)}$ are linear in the Grassmann variables and inherit their property,
\beq
V_{i}^{(p)} V_{j}^{(q)} = (-1)^{\delta^{pq} \delta^{ij}} V_{j}^{(q)} V_{i}^{(p)}\,, \quad\qquad p,q \in \{1, \ldots, \ntheta \}\,,
\label{eq:V_anticommutes}
\eeq
each $V_{i}^{(p)}$ can appear only once in a given term.\footnote{A simple corollary is the well-known fact that two scalar operators 
couple only to fully symmetric representations.} 
Thus, to the $p$-th column of  the irrep $\lambda_i$ there may be only one $V_{i}^{(p)}$, and therefore  there can be at most
$l_1^{i}$ of the $V_{i}^{(p)}$'s in a given tensor structure.
To illustrate this, the boxes of the following Young diagram that may be assigned to $V_{i}^{(p)}$'s
are shaded
\beq
\bt{young_remove} \quad .
\eeq

Now consider the tensor structures that contain $q$ of the $V_{i}^{(p)}$ building blocks.
To the remaining boxes in the Young diagrams we assign a linear combination of the 
$H_{ij}^{(p,q)}$'s, therefore $q$ is even (odd) if the total number of boxes in all diagrams  $ |\lambda_1| + |\lambda_2| + |\lambda_3|$ is
even  (odd). It is also clear that $q$ can take values in the range
\beq
q \in \left\{0, 1, \ldots, l_1^{1} +l_1^{2} +l_1^{3} \right\}\,.
\eeq
The number of  such independent tensor structures,  containing $q$ of the $V_{i}^{(p)}$ building blocks, 
is given by the multiplicity of the scalar representation in the tensor product of the three irreps under consideration and one
Young diagram consisting of one row of length $q$, i.e.\ in the product
\beq
\lambda_1 \otimes \lambda_2 \otimes \lambda_3 \otimes [q] \,.
\eeq
Hence the total number of tensor structures in a three-point function of operators in irreps $\lambda_1$, $\lambda_2$ and $\lambda_3$ is
\beq
\sum\limits_{q=0}^{l_1^{1} +l_1^2 +l_1^3} b_{\lambda_1 \lambda_2 \lambda_3 [q]} \, ,
\label{eq:number_3pt_structures}
\eeq
with $b_{\lambda_1 \lambda_2 \lambda_3 [q]}$ given in \eqref{eq:higher_tp_coeff_SO}. Notice that the term in the sum with $q=0$ counts structures 
made only out of the $H_{ij}^{(p,q)}$'s considered first above.

To prove the result \eqref{eq:number_3pt_structures}, we resort to the correspondence between three-point functions and leading OPE coefficients
established in \cite{Mack:1976pa,Osborn:1993cr} and discussed in the context of the embedding formalism
in \cite{Costa:2011mg}. We start with the leading terms in the OPE of operators $\calO_i$ in arbitary irreps labeled by $[\lambda_i,\Delta_i]$
using physical space coordinates $x_i^a$ and polarizations ${\bf z}_i^a$, following the discussion in \cite{Costa:2011mg}
\beq
\calO_1(x_1,{\bf z}_1)\, \calO_2(x_2,{\bf z}_2) \sim \sum\limits_{k} \calO_k(x_1,{\bf D_{{\bf z}_k}})\,
t(x_{12},{\bf z}_1,{\bf z}_2,{\bf z}_k) \left( x_{12}^2 \right)^{-\frac{\Delta_1 + \Delta_2 - \Delta_k + |\lambda_1| + |\lambda_2| + |\lambda_k|}{2}} .
\label{eq:OPE}
\eeq
When this is inserted into a three-point function $\big\langle \calO_1 \calO_2 \calO_3 \big\rangle$,
only $\calO_k=\calO_3$ contributes.
$t(x_{12},{\bf z}_1,{\bf z}_2,{\bf z}_3)$ is a rotationally invariant polynomial which scales as
\beq
t(\alpha x_{12},\bbeta_1 {\bf z}_1,\bbeta_2{\bf z}_2,\bbeta_3{\bf z}_3)
=
t(x_{12},{\bf z}_1,{\bf z}_2,{\bf z}_3)
\prod\limits_{i=1}^3
 \alpha^{|\lambda_i|}\left( \beta_i^{(1)}\right)^{l_1^i} \ldots  \left( \beta_i^{(\nz)}\right)^{l_{\nz}^i} \left( \beta_i^{(\theta)} \right)^{(\lambda^t_i)_{_1}} ,
\eeq
where we defined
\beq
\bbeta_{i} {\bf z}_i = \left(\beta_i^{(1)}z_i^{(1)},\ldots,\beta_i^{(\nz)}z_i^{(\nz)},\b_i^{(\theta)}\theta_i^{(1)}\right),
\eeq
for arbitrary constants $\beta_i^{(p)}$. The number of independent tensor structures in $\big\langle \calO_1 \calO_2 \calO_3 \big\rangle$
is now equal to the number of strucutures in $t(x_{12},{\bf z}_1,{\bf z}_2,{\bf z}_3)$, which is clearly given by \eqref{eq:number_3pt_structures}.
Note that the sum in \eqref{eq:number_3pt_structures} arises because the combination of vectors $x_{12}^a$ is a symmetric power of the fundamental representation,
which decomposes into traceless symmetric tensors of the same or smaller rank (with the ranks being all even or all odd). 
\beq
\text{Sym}^n\big( \; \btm{young_1} \; \big)
=[n] \oplus [n-2] \oplus [n-4] \oplus \ldots \, .
\eeq
Next we analyze some examples.

\subsubsection{Example: (Two-form)-Vector-Scalar}

We start with a simple example of a two-form, a vector and a scalar, $\lambda_1 = \bts{young_01}\,, \lambda_2 = \bts{young_1}\,$, $\lambda_3 = \bullet\,$.
As already explained for the two-point function of a $p$-form, there is no need to introduce commuting polarizations for the two-form. Also, for the vector, 
there is obviously no need to introduce any symmetrization or antisymmetrization. It has $n_{Z_2} = n_{\Theta_2} = 0$, therefore
one can freely choose whether to use $Z_2$ or $\Theta_2$ as polarization.
In this case the only possible tensor structure has $q=1$, hence there is one $V_i$ building block. This is simple to see,
since
\beq
 \btm{young_01} \otimes  \btm{young_1}\otimes \bullet = 
  \btm{young_1} \oplus  \btm{young_11}  \oplus \btm{young_001}\,,
\eeq
whose product with $[q]$ has a scalar representation only for $q=1$. The corresponding tensor structure gives a 
three-point function  of the form
\begin{align}
&G_{\chi_1 \chi_2 \chi_3} \big(\{P_i\};\Theta_1,Z_2\big)=
\frac{
V_1^{(\Theta)} H_{12}^{(\Theta,Z)}
}
{(P_{12})^{\frac{\Delta_1+\Delta_2-\Delta_3+3}{2}} (P_{23})^{\frac{\Delta_2+\Delta_3-\Delta_1-1}{2}} (P_{31})^{\frac{\Delta_3+\Delta_1-\Delta_2+1}{2}}} 
\label{eq:ex1_three-point}
\\
=\,&\,
\frac{
-4 \Big( (P_2 \cdot \Theta_1)(P_1 \cdot P_3) - (P_2 \cdot P_1)(\Theta_1 \cdot P_3) \Big)
 \Big( (P_2 \cdot \Theta_1) (P_1 \cdot Z_2) - (\Theta_1 \cdot Z_2) (P_1 \cdot P_2)\Big)
}
{(P_{12})^{\frac{\Delta_1+\Delta_2-\Delta_3+3}{2}} (P_{23})^{\frac{\Delta_2+\Delta_3-\Delta_1+1}{2}} (P_{31})^{\frac{\Delta_3+\Delta_1-\Delta_2+1}{2}}} \,.
\nonumber
\end{align}
It is a mechanical computation to act on this polynomial with the derivatives $\partial_{\Theta_1^A}\partial_{\Theta_1^B}\partial_{Z_2^C}$ to obtain the
components of the corresponding tensor in the embedding space.

\subsubsection{Example: Two-form-Vector-Vector}

Next we consider the three-point function of a two-form and two vectors, $\lambda_1 = \bts{young_01}\,$, $\lambda_2 =\lambda_3 = \bts{young_1}\,$.
In this case there are three possible tensor structures, 
\bea
&q = 0 \qquad &\rightarrow \qquad &&H_{12}^{(\Theta,Z)} H_{13}^{(\Theta,Z)}\,, \\
&q = 2 \qquad &\rightarrow \qquad &&V_{1}^{(\Theta)} V_{2}^{(Z)} H_{13}^{(\Theta,Z)} \text{ and } V_{1}^{(\Theta)} V_{3}^{(Z)} H_{12}^{(\Theta,Z)}\,.
\eea{eq:ex2_three-point_structures}
This can be seen from the product
\beq
 \btm{young_01} \otimes  \btm{young_1}\otimes \btm{young_1}= 
 \bullet \oplus 2\,  \btm{young_2} \oplus 3\,\btm{young_01} \oplus \btm{young_21}\oplus \btm{young_02} \oplus 2\, \btm{young_101}
 \oplus \btm{young_0001}\ ,
\eeq
which contains the scalar and $\bts{young_2}$ representations  with multiplicities one and two, respectively.
The corresponding  three-point function has the form
\bea
G_{\chi_1 \chi_2 \chi_3} \big(\{P_i\};\Theta_1,Z_2,Z_3\big)
=
\frac{
c_1 H_{12}^{(\Theta,Z)} H_{13}^{(\Theta,Z)}
+c_2 V_{1}^{(\Theta)} V_{2}^{(Z)} H_{13}^{(\Theta,Z)}
+c_3 V_{1}^{(\Theta)} V_{3}^{(Z)} H_{12}^{(\Theta,Z)}
}
{(P_{12})^{\frac{\Delta_1+\Delta_2-\Delta_3+2}{2}} (P_{23})^{\frac{\Delta_2+\Delta_3-\Delta_1}{2}} (P_{31})^{\frac{\Delta_3+\Delta_1-\Delta_2+2}{2}}} \,,
\eea{eq:ex2_three-point}
with $c_1$, $c_2$ and $c_3$ constants.

\subsubsection{Example: Hook-Scalar-Vector}

The polynomial that encodes the correlator of a small hook diagram $\lambda_1 = \bts{young_11}\,$, a scalar $\lambda_2 = \bullet$ and a vector $\lambda_3 = \bts{young_1}\,$
consists of a single tensor structure, as can easily seen by considering the product
\beq
\btm{young_11}  \otimes \bullet \otimes \btm{young_1}= 
 \btm{young_2} \oplus \btm{young_01} \oplus \btm{young_21}\oplus \btm{young_02} \oplus  \btm{young_101}\,.
\eeq
Recall that for the small hook diagram we have $n_{Z_1} = 1$ and $n_{\Theta_1} = 1$, with polarization vectors  ${\bf Z}_1=(Z_1,\Theta_1)$ and $\bTheta_1=(\Theta_1,Z_1)$,
so the tensor structure is obtained by acting with a derivative $Z_1 \cdot \partial_{\Theta_1}$ on a polynomial of 
the $V_{i}^{(p)}$'s and $H_{ij}^{(p,q)}$'s. In this case the single tensor structure has the form
\bea
G_{\chi_1 \chi_2 \chi_3} \big(\{P_i\} ; {\bf Z}_1,Z_3\big)
={}&
\frac{\left( Z_1 \cdot \partial_{\Theta_1} \right)  V_1^{(\Theta)} V_1^{(Z)} H_{13}^{(\Theta,Z)}}
{(P_{12})^{\frac{\Delta_1+\Delta_2-\Delta_3+2}{2}} (P_{23})^{\frac{\Delta_2+\Delta_3-\Delta_1-2}{2}} (P_{31})^{\frac{\Delta_3+\Delta_1-\Delta_2+4}{2}}}\\
={}& \frac{V_1^{(\Theta)} V_1^{(Z)} H_{13}^{(Z,Z)} - \left( V_1^{(Z)}\right)^2 H_{13}^{(\Theta,Z)}}
{(P_{12})^{\frac{\Delta_1+\Delta_2-\Delta_3+2}{2}} (P_{23})^{\frac{\Delta_2+\Delta_3-\Delta_1-2}{2}} (P_{31})^{\frac{\Delta_3+\Delta_1-\Delta_2+4}{2}}} \,.
\eea{eq:3pt_hook_scalar_vector}

\subsubsection{Example: Hook-Spin 2-Vector}
\label{sec:example_hook_spin2_vector}

Let us finally consider the example 
 $\lambda_1 = \bts{young_11}\,$, $\lambda_2 = \bts{young_2}\,$, $\lambda_3 = \bts{young_1}\,$.
Table \ref{table:3pt:example} contains all independent tensor structures for this case.
Notice that for $q=2$ there is another tensor structure constructed from $V_1^{(Z)}V_2^{(Z)}H_{12}^{(\Theta,Z)}H_{13}^{(\Theta,Z)}$,
but this is not linear independent since
\bea
\left( Z_1 \cdot \partial_{\Theta_1} \right) & V_1^{(Z)}V_2^{(Z)}H_{12}^{(\Theta,Z)}H_{13}^{(\Theta,Z)} \\
&= 
\left( Z_1 \cdot \partial_{\Theta_1} \right) 
\left(
V_1^{(\Theta)}V_2^{(Z)}H_{12}^{(Z,Z)}H_{13}^{(\Theta,Z)} -
V_1^{(\Theta)}V_2^{(Z)}H_{12}^{(\Theta,Z)}H_{13}^{(Z,Z)} 
\right)
\,.
\eea{eq:ex_linear_dep}
In this case the product of the three representations $\lambda_1$, $\lambda_2$ and $\lambda_3$ 
contains the following representations consisting of a single row
\beq
\btm{young_11}  \otimes \btm{young_2} \otimes \btm{young_1}= \bullet \oplus 4\,  \btm{young_2} \oplus  2\,\btm{young_4}
\oplus \ldots\,,
\eeq
in agreement with Table \ref{table:3pt:example}.

\begin{table}[t]
\centering
{\renewcommand{\arraystretch}{1.3}
\begin{tabular}{c|c|l}
$q$ &  $b_{\lambda_1 \lambda_2 \lambda_3 [q]}$ &tensor structures  \\ \hline
0 & 1&
$H_{12}^{(Z,Z)}  H_{12}^{(\Theta,Z)} H_{13}^{(\Theta,Z)}$ \\
2 & 4&
$V_1^{(\Theta)} V_1^{(Z)} H_{12}^{(\Theta,Z)}  H_{23}^{(Z,Z)}$,
$V_1^{(\Theta)} V_3^{(Z)} H_{12}^{(Z,Z)}  H_{12}^{(\Theta,Z)}$, \\  
 & &
$V_1^{(\Theta)} V_2^{(Z)} H_{12}^{(Z,Z)}  H_{13}^{(\Theta,Z)}$,
$V_1^{(\Theta)} V_2^{(Z)} H_{12}^{(\Theta,Z)}  H_{13}^{(Z,Z)}$\\
4 & 2&
$V_1^{(\Theta)} V_1^{(Z)}V_2^{(Z)}V_3^{(Z)} H_{12}^{(\Theta,Z)}$,
$V_1^{(\Theta)} V_1^{(Z)}(V_2^{(Z)})^2 H_{13}^{(\Theta,Z)}$ \\
\end{tabular}
}
\caption[]{All seven tensor structures appearing in a three-point function of irreps $\bts{young_11}\,$, $\bts{young_2}$ and $\bts{young_1}\,$.}
\label{table:3pt:example}
\end{table}

\subsection{Four-point functions}

Starting from four-point functions, correlation functions can depend on functions of the conformally invariant
cross-rations. For four points there are two cross-ratios that can be defined to be
\beq
u = \frac{P_{12} P_{34}}{P_{13} P_{24}}\,, \qquad
v = \frac{P_{14} P_{23}}{P_{13} P_{24}}\,.
\label{eq:cross_ratios}
\eeq
Then a  generic four-point function can be written as
\begin{align}
G_{\chi_1 \chi_2 \chi_3 \chi_4} \big(\{P_i;\bZ_i\}\big)= &\ 
\frac{\left( \frac{P_{24}}{P_{14}} \right)^{\frac{\tau_1-\tau_2}{2}}  \left( \frac{P_{14}}{P_{13}} \right)^{\frac{\tau_3-\tau_4}{2}}}
{\left( P_{12} \right)^{\frac{\tau_1+\tau_2}{2}}  \left( P_{34} \right)^{\frac{\tau_3+\tau_4}{2}}}\times
\label{eq:def_four-point} 
\\
&
\prod\limits_{j=1}^{4}
\prod\limits_{p=1}^{\nz^j} \prod\limits_{q=1}^{\min ( l_p^j,\ntheta^j)}
\left( Z^{(p)}_j \cdot \partial_{\Theta^{(q)}_j} \right) 
\sum\limits_k f_k(u,v) \,\bar Q_{\chi_1 \chi_2 \chi_3 \chi_4}^{(k)} \big(\{P_i;\bTheta_i\}\big) \,,
\nonumber
\end{align}
where $\tau_i = \Delta_i + |\lambda_i|$ and the sum over $k$ runs over all independent tensor structures. Each
tensor structure is multiplied by a function of the cross-ratios $f_k(u,v)$ and the pre-factor is chosen in such a way that each
$\bar Q_{\chi_1\chi_2 \chi_3 \chi_4}^{(k)}$ scales analogously to \eqref{eq:3point-scale},
\bea
&\bar Q_{\chi_1 \chi_2 \chi_3 \chi_4}^{(k)} \big(\{\alpha_i P_i; \bbeta_{i} (\bTheta_i+ \bgamma_{i}P_i)\}\big)=\\
={}&\bar Q_{\chi_1 \chi_2 \chi_3 \chi_4}^{(k)} \big(\{P_i;\bTheta_i\}\big) 
\prod\limits_i \alpha_i^{|\lambda_i|} 
\left(\beta_i^{(1)}\right)^{h^i_1} \ldots \left(\beta_i^{(\ntheta^i)} \right)^{h^i_{\ntheta^i}} \left( \beta_i^{(Z)}\right)^{(\lambda_i)_{_1}} 
 \,.
\eea{eq:4point-scale}
For an $n$-point function, $n-2$ of the building blocks $V^{(p)}_{i,jk}$ are linearly independent \cite{Costa:2011mg}.
In the case of $n=4$, this is due to the relations $V^{(p)}_{i,jk} = -V^{(p)}_{i,kj}$ and
\beq
\big(P_2\cdot P_3\big)\big(P_1\cdot P_4\big) V^{(p)}_{1,23}
+ \big(P_3\cdot P_4\big)\big(P_1\cdot P_2\big) V^{(p)}_{1,34}
+ \big(P_4\cdot P_2\big)\big(P_1\cdot P_3\big) V^{(p)}_{1,42}
=0 \,.
\eeq
In general one can choose for instance the basis
\beq
\mathcal{V}^{(p)}_{ij} \equiv V^{(p)}_{i,(i+1)(i+1+j)}\,,  \qquad j \in \{1, 2, \ldots, n-2 \} \,,
\label{eq:def_cal_V}
\eeq
where the external point labels $i$ etc.\ are meant to be interpreted modulo $n$.

The tensor structures can be counted by inserting the OPE \eqref{eq:OPE} twice into the four-point function
\begin{align}
& \calO_1(x_1,{\bf z}_1) \,\calO_2(x_2,{\bf z}_2) \,\calO_3(x_3,{\bf z}_3) \nonumber\\
\sim{}& \calO_1(x_1,{\bf z}_1) \sum\limits_{k} \calO_k(x_2,{\bf D_{{\bf z}_k}})\,
t(x_{23},{\bf z}_2,{\bf z}_3,{\bf z}_k) \left( x_{23}^2 \right)^{-\frac{\Delta_2 + \Delta_3 - \Delta_k + |\lambda_2|+|\lambda_3|+|\lambda_k| }{2}} \label{eq:OPE_4pt}\\
\sim{}& 
\sum\limits_{j} \calO_j(x_1, {\bf D_{{\bf z}_j}})
\sum\limits_{k} 
\frac{
t(x_{12},{\bf z}_1, {\bf D_{{\bf z}_k}},{\bf z}_j)\,
t(x_{23},{\bf z}_2,{\bf z}_3,{\bf z}_k) }{
\left( x_{12}^2 \right)^{\frac{\Delta_1 + \Delta_k - \Delta_j + |\lambda_1|+|\lambda_k|+|\lambda_j|}{2}} \left( x_{23}^2 \right)^{\frac{\Delta_2 + \Delta_3 - \Delta_k + |\lambda_2|+|\lambda_3|+|\lambda_k|}{2}}
}
\,.\nonumber
\end{align}
When this is inserted into $\big\langle \calO_1 \calO_2 \calO_3 \calO_4 \big\rangle$, only $\calO_j=\calO_4$ contributes.
When summed over all possible irreps $k$, the terms 
$t(x_{12},{\bf z}_1, {\bf D_{{\bf z}_k}},{\bf z}_4)\,t(x_{23},{\bf z}_2,{\bf z}_3,{\bf z}_k)$ 
clearly contain all possible contractions of 
$x_{12}$, $x_{23}$ and the four polarizations ${\bf z}_1$, ${\bf z}_2$, ${\bf z}_3$ and ${\bf z}_4$.
To exclude contractions between $x_{12}$ and $x_{23}$, which do not lead to new tensor structures,
counting can be performed using the restricted tensor product defined in \eqref{eq:square_tensor_product}, 
that keeps only irreps that have the same number of indices as both irreps in the product.
The number of tensor structures in a four-point function are then given by the multiplicity of the scalar
representation in the tensor product
\beq
\lambda_1 \otimes \lambda_2 \otimes \lambda_3 \otimes \lambda_4 \otimes \Big[ [q_1] \otimes [q_2]\Big] \,,
\label{eq:tensor4ptstructures}
\eeq
for non-negative integers  $q_1$ and $q_2$ satisfying 
\beq
q_1, q_2 \in \Big\{0, 1, \ldots, \sum\limits_{i=1}^4 l_1^i \Big\} \, .
\eeq
Hence the total number of tensor structures is
\beq
\sum\limits_{q_1=0}^{\sum_i l_1^i}
\sum\limits_{q_2=0}^{\sum_i l_1^i}
\sum\limits_\mu c_{[q_1] [q_2]}^{\qquad \ \mu} \,  b_{\mu \lambda_1 \lambda_2 \lambda_3 \lambda_4} 
\equiv \sum\limits_{q_1=0}^{\sum_i l_1^i} \sum\limits_{q_2=0}^{\sum_i l_1^i} d(q_1,q_2)
\, ,
\label{eq:number_4pt_structures}
\eeq
where in the last equality we defined the number of structures $d(q_1,q_2)$ for a given pair $(q_1,q_2)$. In fact,  
when constructing the tensor structures, it is helpful to treat the contributions for each combination $(q_1,q_2)$
separately. We make the assignment that the tensor structures corresponding to $(q_1,q_2)$
contain $q_1$ building blocks $\mathcal{V}^{(p)}_{i1}$ and $q_2$ building blocks  $\mathcal{V}^{(p)}_{i2}$.
To make sense of this assignment one can consider two boxes in the same column $p$ of one of the Young diagrams $\lambda_i$.
The only way to assign these two boxes to $\mathcal{V}^{(p)}_{ij}$ building blocks is to assign one to $\mathcal{V}^{(p)}_{i1}$
and one to $\mathcal{V}^{(p)}_{i2}$, because of \eqref{eq:V_anticommutes}.
Correspondingly, in the decomposition of the tensor product of $[q_1] \otimes [q_2]$ vertically aligned boxes always consist of one box belonging to $[q_1]$ and one to $[q_2]$. Finally, notice that for a given pair $(q_1,q_2)$ there may be more than 
$d(q_1,q_2)$ possible combinations of building blocks $H_{ij}^{(p,q)}$ and $\mathcal{V}^{(p)}_{ij}$
that contain $q_1$ building blocks $\mathcal{V}^{(p)}_{i1}$ and $q_2$  building blocks $\mathcal{V}^{(p)}_{i2}$,
however only $d(q_1,q_2)$ of these combinations will give raise to linearly independent structures after antisymmetrization
(just as in the example of section \ref{sec:example_hook_spin2_vector}).

\subsubsection{Example: Scalar-Vector-Scalar-Vector}
\label{sec:example_sc-v-sc-v}

As an example, Table \ref{table:4pt_svsv_example} lists the five tensor structures
in a four-point function of irreps $\lambda_1 = \lambda_3 = \bullet\,$ and
$\lambda_2 = \lambda_4 = \bts{young_1}\,$, which were already given in  \cite{Cornalba:2009ax,Costa:2011mg}.
Let us see explicitly how these structures arise from the scalar degeneracy in the tensor product \eqref{eq:tensor4ptstructures}.
The tensor product of  two scalars   and two vectors decomposes as
\beq
\bullet \otimes\, \btm{young_1} \otimes\bullet \otimes \btm{young_1} = \btm{young_1} \otimes \btm{young_1}
=  \bullet\oplus\,  \btm{young_2} \oplus \btm{young_01} \, .
\eeq
That the scalar representation appears with multiplicity one here means that
there is one tensor structure for $(q_1,q_2) = (0,0)$, i.e. $d(0,0)=1$. For $(q_1,q_2) = (2,0)$ or $(q_1,q_2) = (0,2)$
we have to consider the tensor product
\beq
\bullet \otimes\, \btm{young_1} \otimes\bullet \otimes \btm{young_1}  \otimes \btm{young_2}
=
\left( \bullet\oplus\,  \btm{young_2} \oplus \btm{young_01} \ \right) \otimes \btm{young_2} = \bullet \oplus \text{other irreps}\,.
\eeq
Thus $d(2,0)=d(0,2)=1$.
Finally, for $(q_1,q_2) = (1,1)$ one needs to consider
\beq
\bullet\,\otimes\, \btm{young_1}\, \otimes\,\bullet \,\otimes\, \btm{young_1}\, \otimes 
\Big[ \ \btm{young_1} \; \otimes \; \btm{young_1} \ \Big]
=
\left( \bullet\oplus\,  \btm{young_2} \oplus \btm{young_01} \ \right) \otimes
\left( \ 
\btm{young_2} \oplus \btm{young_01} 
\ \right)
= 2 \bullet \,\oplus \text{ other irreps}\,,
\eeq
so that $d(1,1)=2$.

\begin{table}[h]
\centering
{\renewcommand{\arraystretch}{1.3}
\begin{tabular}{cc|c|l}
$q_1$ & $q_2$ & $d(q_1,q_2)$ & tensor structures \\ \hline
0 & 0  & 1 & $H_{24}^{(Z,Z)}$ \\
2 & 0  & 1 & $\mathcal{V}^{(Z)}_{21} \mathcal{V}^{(Z)}_{41}$ \\
1 & 1  & 2 & $\mathcal{V}^{(Z)}_{21} \mathcal{V}^{(Z)}_{42}$, $\mathcal{V}^{(Z)}_{22} \mathcal{V}^{(Z)}_{41}$ \\
0 & 2  & 1 & $\mathcal{V}^{(Z)}_{22} \mathcal{V}^{(Z)}_{42}$
\end{tabular}
}
\caption[]{All five tensor structures in a four-point function of irreps $\bullet\,$, $\bts{young_1}\,$, $\bullet\,$, $\bts{young_1}\,$.}
\label{table:4pt_svsv_example}
\end{table}

\subsubsection{Example: Hook-Vector-Scalar-Scalar}

For this example we consider the irreps 
$\lambda_1 = \bts{young_11}\,$, $\lambda_2 = \bts{young_1}\,$, $\lambda_3 = \lambda_4 = \bullet\,$.
Table \ref{table:4pt_hook_example} shows all the eight tensor structures for this correlator.
 This counting can be confirmed by looking at the OPE in the channel
${\cal O}_3\times {\cal O}_4 \sim {\cal O}_l$. Then the number of possible structures in a three-point function of 
$\lambda_1 = \bts{young_11}\,$, $\lambda_2 = \bts{young_1}\,$ and a spin $l$ symmetric traceless tensor is also seen to be
eight\footnote{This is a simple generalization of the example in subsection \ref{sec:example_hook_spin2_vector} which has $l=2$.
For $l \ge 3$, the three-point function of 
$\lambda_1 = \bts{young_11}\,$, $\lambda_2 = \bts{young_hook0}\,$ with $|\lambda_2|=l$ and $\lambda_3 = \bts{young_1}\,$
has the structures listed in Table \ref{table:3pt:example} each multiplied by $\left(V_2^{(Z)}\right)^{l-2}$
and the
additional structure $H_{12}^{(\Theta,Z)}  H_{12}^{(Z,Z)} H_{23}^{(Z,Z)}  V_1^{(\Theta)} \left(V_2^{(Z)}\right)^{l-3}$.}.

\begin{table}[h]
\centering
{\renewcommand{\arraystretch}{1.3}
\begin{tabular}{cc|l}
$q_1$ & $q_2$ &  tensor structures \\ \hline
2 & 0 & $\mathcal{V}^{(\Theta)}_{11} \mathcal{V}^{(Z)}_{11} H^{(\Theta,Z)}_{12}$\\
1 & 1 & $\mathcal{V}^{(Z)}_{11} \mathcal{V}^{(\Theta)}_{12} H^{(\Theta,Z)}_{12}$, $\mathcal{V}^{(\Theta)}_{11} \mathcal{V}^{(Z)}_{12} H_{12}^{(\Theta,Z)}$\\
0 & 2 & $\mathcal{V}^{(\Theta)}_{12} \mathcal{V}^{(Z)}_{12} H^{(\Theta,Z)}_{12}$\\
3 & 1 & $\mathcal{V}^{(\Theta)}_{11} \mathcal{V}^{(Z)}_{11} \mathcal{V}^{(Z)}_{21} \mathcal{V}^{(\Theta)}_{12}$\\
2 & 2 & $\mathcal{V}^{(\Theta)}_{11} \mathcal{V}^{(Z)}_{11} \mathcal{V}^{(\Theta)}_{12} \mathcal{V}^{(Z)}_{22}$,
$\mathcal{V}^{(\Theta)}_{11} \mathcal{V}^{(Z)}_{21} \mathcal{V}^{(\Theta)}_{12} \mathcal{V}^{(Z)}_{12}$\\
1 & 3 & $\mathcal{V}^{(\Theta)}_{11} \mathcal{V}^{(\Theta)}_{12} \mathcal{V}^{(Z)}_{12} \mathcal{V}^{(Z)}_{22}$
\end{tabular}
}
\caption[]{All eight tensor structures in a four-point function of irreps $\bts{young_11}\,$, $\bts{young_1}\,$, $\bullet\,$, $\bullet\,$.}
\label{table:4pt_hook_example}
\end{table}

Let us see again explicitly how these structures arise from the scalar degeneracy in the tensor product \eqref{eq:tensor4ptstructures}.
First we consider the product
\beq
 \btm{young_11}  \otimes  \btm{young_1}\otimes  \bullet\otimes  \bullet
 =  \btm{young_2} \oplus \btm{young_01}\oplus \btm{young_21} \oplus \btm{young_02} \oplus  \btm{young_101}  \,.
 \label{eq:4pt_hook_example}
\eeq
Since there is no scalar irrep in this sum we have $d(0,0)=0$. For the other values of $(q_1,q_2)$, \eqref{eq:4pt_hook_example} 
must be multiplied by $\big[ [q_1] \otimes [q_2]\big]$, and then $d(q_1,q_2)$ is just the multiplicity of the scalar irrep in 
the overall product. Table \ref{table:4pt_hook_q1q2products} shows the different possibilities.
\begin{table}[h]
\centering
{\renewcommand{\arraystretch}{1.6}
\begin{tabular}{cc|c|c}
 $q_1$ & $q_2$ & $\big[ [q_1] \otimes [q_2]\big]$ & $d(q_1,q_2)$\\ \hline
 2 & 0 & $\btm{young_2}$ & 1 \\
1 & 1 &  $\btm{young_2} \oplus \btm{young_01}$ & 2\\
0 & 2 &  $\btm{young_2}$ & 1\\
3 & 1 & $\btm{young_4} \oplus \btm{young_21}$ & 1 \\
2 & 2 & $\btm{young_4}  \oplus \btm{young_21}\oplus \btm{young_02}$ &2 \\
1 & 3 &  $\btm{young_4} \oplus \btm{young_21}$ & 1 
\end{tabular}
}
\caption[]{From the product of $\big[ [q_1] \otimes [q_2]\big]$ with \eqref{eq:4pt_hook_example} it is straightforward to extract the 
scalar multiplicity $d(q_1,q_2)$, which counts the independent  tensor structures given in Table \ref{table:4pt_hook_example}.}
\label{table:4pt_hook_q1q2products}
\end{table}

\subsubsection{Example: Vector-Vector-Vector-Vector}

Finally, the correlation function of four vectors illustrates how the tensor product also generates the number
of possible contractions between $H$'s, i.e. those corresponding to 
$q_1 = q_2 =0$. The number of such
tensor structures is calculated using the $SO(d)$ tensor product
\bea
\btm{young_1} \otimes \btm{young_1} \otimes \btm{young_1} \otimes \btm{young_1}
= 3 \bullet\, \oplus \,6\, \btm{young_2} \oplus 6\,\btm{young_01} \oplus \btm{young_4} \oplus 3\,\btm{young_21}\oplus2\,\btm{young_02}\oplus3\,\btm{young_101} \oplus \btm{young_0001}\ .
\eea{eq:4pt_vector_example}
Correspondingly, there are three tensor structures that can be built out of $H$'s, namely
\beq
H_{12}^{(Z,Z)} H_{34}^{(Z,Z)} , \quad H_{13}^{(Z,Z)} H_{24}^{(Z,Z)} 
\quad \text{and} \quad H_{14}^{(Z,Z)} H_{23}^{(Z,Z)} \,.
\eeq
There are $3! 2^2$ other structures with two $\mathcal{V}$'s and one $H$  and $2^4$ other structures with four $\mathcal{V}$'s. 
Thus, in total for this case there are 43 independent tensor structures.  
As in the previous example, this counting is done by considering the 
scalar multiplicity in the product of $\big[ [q_1] \otimes [q_2]\big]$ with \eqref{eq:4pt_vector_example}.
Table \ref{table:4pt_vector_q1q2products} shows the different possibilities to which it is trivial to assign the
independent tensor structures.
\begin{table}[h]
\centering
{\renewcommand{\arraystretch}{1.6}
\begin{tabular}{cc|c|c}
 $q_1$ & $q_2$ & $\big[ [q_1] \otimes [q_2]\big]$ & $d(q_1,q_2)$\\ \hline
 0 & 0 & $\bullet$ & 3 \\
 2 & 0 & $\btm{young_2}$ & 6 \\
1 & 1 &  $\btm{young_2} \oplus \btm{young_01}$ & 12\\
0 & 2 &  $\btm{young_2}$ & 6\\
4 & 0 & $\btm{young_4}$ & 1 \\
3 & 1 & $\btm{young_4} \oplus \btm{young_21}$ & 4 \\
2 & 2 & $\btm{young_4} \oplus\btm{young_21}\oplus\btm{young_02}$ &6 \\
1 & 3 &  $\btm{young_4}\oplus \btm{young_21}$ & 4 \\
0 & 4 & $\btm{young_4}$ & 1 
\end{tabular}
}
\caption[]{Multiplicity $d(q_1,q_2)$ counting  tensor structures for the correlation function of four vectors.}
\label{table:4pt_vector_q1q2products}
\end{table}

\subsection{$n$-point functions}

Let us comment briefly on the general construction of $n$-point functions. It is analogous to the construction of four-point functions.
Generically one can write,
\bea
G_{\chi_1 \ldots \chi_n} \big(\{P_i;\bZ_i\}\big)
=
\prod\limits_{g<h}^n P_{gh}^{-\alpha_{gh}}
\prod\limits_{j=1}^{n}
\prod\limits_{p=1}^{\nz^j} \prod\limits_{q=1}^{\min ( l_p^j,\ntheta^j)}
\!\!\left( Z^{(p)}_j \cdot \partial_{\Theta^{(q)}_j} \right) 
& \sum\limits_k f_k(u_a) \,\bar Q_{\chi_1\ldots \chi_n}^{(k)} \big(\{P_i;\bTheta_i\}\big) \,,
\eea{eq:def_n-point} 
where $u_a$ are the $n(n-3)/2$ independent conformally invariant cross-ratios,
\beq
\alpha_{gh} = \frac{\tau_g + \tau_h}{n-2} - \frac{1}{(n-1)(n-2)} \sum\limits_{i=1}^n \tau_i \,,
\eeq
and the pre-factor is  chosen to let the functions $\bar Q_{\chi_1\ldots \chi_n}^{(k)}$ scale in the already familiar way
\bea
&\bar Q_{\chi_1 \ldots \chi_n}^{(k)} \big(\{\alpha_i P_i; \bbeta_{i}  (\bTheta_i+ \bgamma_{i}P_i)\}\big)=\\
={}&\bar Q_{\chi_1 \ldots \chi_n}^{(k)} \big(\{P_i;\bTheta_i\}\big) 
\prod\limits_i \alpha_i^{|\lambda_i|} 
\left(\beta_i^{(1)}\right)^{h^i_1} \ldots \left(\beta_i^{(\ntheta^i)} \right)^{h^i_{\ntheta^i}} \left( \beta_i^{(Z)}\right)^{(\lambda_i)_1} 
\,.
\eea{eq:npoint-scale}
These functions are again constructed from the building blocks $H^{(p)}_{ij}$ and $\mathcal{V}^{(p)}_{ij}$ defined in \eqref{eq:Hij_def} and \eqref{eq:def_cal_V}.
The counting of tensor structures is done as described for four-point functions in the previous section,
but now the tensor product contains $n-2$ additional representations for counting all the combinations of $\mathcal{V}^{(p)}_{ij}$ building blocks,
since there are that many independent  $\mathcal{V}^{(p)}_{ij}$ for each $i$.
The resulting number of tensor structures in a correlator of $n$ operators in the irreps $\chi_i = [\lambda_i, \Delta_i]$ is 
the multiplicity of the scalar representation in the product 
\beq
\lambda_1 \otimes \ldots \otimes \lambda_n \otimes \Big[ [q_1] \otimes \ldots \otimes [q_{n-2}]\Big]  \,,
\eeq
which is given by
\beq
\sum\limits_{q_1,\ldots, q_{n-2}} 
\,\,
\sum\limits_\mu c_{[q_1] \ldots [q_{n-2}]}^{\qquad \qquad \mu} \,  b_{\mu \lambda_1 \ldots \lambda_n} \, ,
\label{eq:number_npt_structures}
\eeq
where the sums run over non-negative $q_j$'s with
\beq
q_j \in  \Big\{0, 1, \ldots, \sum\limits_{i=1}^n l_1^i \Big\} \, .
\eeq
As for four-point functions, to construct the tensor structures it is helpful 
to assign to each  $(q_1, \ldots, q_{n-2})$
 the tensor structures with $q_1$ copies of $\mathcal{V}^{(p)}_{i1}$,
$q_2$ copies of $\mathcal{V}^{(p)}_{i2}$, and so on.

\section{Conserved tensors}
\label{sec:conserved_tensors}

Let us now consider conserved tensors in arbitrary irreducible $SO(d)$ representations.
Recall that the unitarity bound for mixed-symmetry tensors  \cite{Metsaev:1995re,Minwalla:1997ka},
that must be satisfied in unitary CFTs, 
restricts the conformal dimension of primaries in the irrep $\lambda$ to satisfy the condition
\beq
\Delta \geq l^\lambda_1 - h^\lambda_{l_1} + d - 1\,,
\label{eq:unitarity_bound}
\eeq
where $h^\lambda_{l_1}$ is the height of the rightmost column (the number of upper rows with the same number of boxes).
The dimension for which \eqref{eq:unitarity_bound} is saturated is called the critical dimension.

Let us first recall that, at the critical dimension, the conservation condition on
fully symmetric or fully antisymmetric tensors $f_{a_1\ldots a_{l}} (x)$,
\beq
\frac{\partial}{\partial x_{a_1}} f_{a_1\ldots a_{l}} (x) = 0 \,,
\label{eq:conservation_general}
\eeq
is conformally invariant.
The question which equations are conformally invariant for more general representations
of the conformal group was discussed in \cite{Shaynkman:2004vu}, and specifically for mixed-symmetry
tensors of hook diagram type in \cite{Alkalaev:2012rg}.
We will show below that, for general mixed-symmetry tensors in irrep $\lambda$, the analogue of the conservation condition
\eqref{eq:conservation_general} can only be imposed with respect to indices that correspond to boxes in one of the lowest columns
in the Young tableau, i.e.\ they can be written as
\beq
\frac{\partial}{\partial x_{g_1}} f_{[a_1\ldots a_{h_1}][b_1\ldots b_{h_2}]\ldots[g_1\ldots g_{h_{l_1}}]} (x) = 0 \,.
\label{eq:conservation}
\eeq
We will see that this equation can be imposed directly in embedding space.
At the same time this will allow us to see that it is conformally invariant
only when the unitarity bound \eqref{eq:unitarity_bound} is saturated,
and that similar equations with the derivative contracted with a different index are not conformally invariant.

The computation was done in \cite{Costa:2011mg} for symmetric tensors and
the only part that changes is when the index symmetries are used.
Let us first write 
\bea
\frac{\partial}{\partial x_{a_{|\lambda|}}} f_{a_1\ldots a_{|\lambda|}}(x)
=&\,\frac{\partial}{\partial x_{a_{|\lambda|}}}  
\left(\frac{\partial P^{A_1}}{\partial x^{a_1}} \ldots \frac{\partial P^{A_{|\lambda|}}}{\partial x^{a_{|\lambda|}}}
F_{A_1 \ldots A_{|\lambda|}} (P_x)  \right) \\
=&\,\frac{\partial P^{A_1}}{\partial x^{a_1}} \ldots \frac{\partial P^{A_{|\lambda|-1}}}{\partial x^{a_{|\lambda|-1}}}
S_{A_1 \ldots A_{|\lambda|-1}} (P_x)
+ T_{a_1 \ldots a_{|\lambda|-1}} (x) \,,
\eea{eq:conserved_derivation_1}
where the projection from $F_{A_1 \ldots A_{|\lambda|}}$ to $f_{a_1\ldots a_{|\lambda|}}$ given in  \eqref{eq:physical_embedding_proj} was inserted and
\bea
S_{A_1 \ldots A_{|\lambda|-1}} (P) ={}&
\left[ \frac{\partial}{\partial P_{A_{|\lambda|}}} - \frac{1}{P \cdot \bar P} \left(\bar P \cdot \frac{\partial}{\partial P} \right)P^{A_{|\lambda|}} 
-(d-1-\Delta) \frac{ \bar P^{A_{|\lambda|}}}{P \cdot \bar P} \right] F_{A_1 \ldots A_{|\lambda|}} (P) \,,
\eea{eq:R_tilde}
is obtained in the same way as in \cite{Costa:2011mg},  with $\bar P = (0,2,0)$ in the light-cone coordinates 
introduced in  \eqref{eq:Poincare}.
The part $T_{a_1 \ldots a_{|\lambda|-1}} (x)$ comprises terms where $\frac{\partial}{\partial x_{a}}$ 
acts on the $\frac{\partial P^{A}}{\partial x^{b}}$ and can be simplified using
\beq
\frac{\partial}{\partial x_{a}} \frac{\partial P^{A}}{\partial x^{b}} = \delta_{ab} \bar P^A .
\eeq 
This is the part where the index symmetries are important
\bea
T_{a_1 \ldots a_{|\lambda|-1}} (x) ={}&
 - \frac{1}{P \cdot \bar P} \frac{\partial P^{A_{|\lambda|}}}{\partial x^{a_{|\lambda|}}} \Bigg[ 
\delta_{a_{|\lambda|} a_1} \bar P^{A_1} \frac{\partial P^{A_2}}{\partial x^{a_2}} \ldots \frac{\partial P^{A_{|\lambda|-1}}}{\partial x^{a_{|\lambda|-1}}} + \ldots \\
&\qquad \qquad \qquad \quad
+
\frac{\partial P^{A_1}}{\partial x^{a_1}}  \ldots \frac{\partial P^{A_{|\lambda|-2}}}{\partial x^{a_{|\lambda|-2}}} \,\delta_{a_{|\lambda|} a_{|\lambda|-1}} \bar P^{A_{|\lambda|-1}}
\Bigg] F_{A_1 \ldots A_{|\lambda|}} (P) \\
={}& - \frac{1}{P \cdot \bar P} \frac{\partial P^{A_1}}{\partial x^{a_1}}
\ldots \frac{\partial P^{A_{|\lambda|-1}}}{\partial x^{a_{|\lambda|-1}}}
\bar P^{A_{|\lambda|}}\\
&\Big[ 
F_{A_{|\lambda|} A_2 \ldots A_{|\lambda|-1} A_1}
+ F_{A_ 1 A_{|\lambda|} A_3 \ldots A_{|\lambda|-1} A_2}
+ \ldots
+ F_{A_ 1 \ldots A_{|\lambda|-2} A_{|\lambda|} A_{|\lambda|-1}}
\Big] \,.
\eea{eq:def_r_tensor}
The second identity here is just a relabelling of indices.
The sum in the last brackets simplifies due to the index symmetries \eqref{eq:antisymmetrization_ex} and becomes
\beq
\big( (l_1-1) - (h_{l_1}-1) \big) F_{A_1 \ldots A_{|\lambda|}}\,.
\label{eq:symmetries_used}
\eeq
Note that this step is only possible since the derivative in \eqref{eq:conservation}
is contracted with an index in the rightmost column of the Young tableau. 
The lift of the conservation condition with respect to the last index is then
\beq
0 = \frac{\partial}{\partial x_{a_{|\lambda|}}} f_{a_1\ldots a_{|\lambda|}}(x)
= \frac{\partial P^{A_1}}{\partial x^{a_1}} \ldots \frac{\partial P^{A_{|\lambda|-1}}}{\partial x^{a_{|\lambda|-1}}}
R_{A_1 \ldots A_{|\lambda|-1}} (P_x) \,,
\label{eq:conservation_last_index}
\eeq
where
\beq
R_{A_1 \mathinner{\ldotp\ldotp} A_{|\lambda|-1}}  \!(P)= 
\!
\left[ \frac{\partial}{\partial P_{A_{|\lambda|}}} - \frac{1}{P \cdot \bar P}\! 
\left(\bar P \cdot \frac{\partial}{\partial P} \right)\!P^{A_{|\lambda|}} 
-\big(l_1 - h_{l_1} +d-1-\Delta\big) \frac{\bar P^{A_{|\lambda|}}}{P \cdot \bar P} \right] 
\!
F_{A_1 \mathinner{\ldotp\ldotp} A_{|\lambda|}} (P).
\label{eq:R_last_index}
\eeq
This generalises the result derived in  \cite{Costa:2011mg} for symmetric tensors. 
As discussed in \cite{Costa:2011mg}, the first two terms in \eqref{eq:R_last_index}
are $SO(d+1,1)$ invariant. The last term is not, but it vanishes for conserved tensors
which saturate the unitarity bound \eqref{eq:unitarity_bound}.
Because of the index symmetries \eqref{eq:index_groups_sym} the derivative in the conservation condition
\eqref{eq:conservation_last_index} can be contracted with any
index that belongs to a column in the Young diagram of the same height as the rightmost one.
In particular it may be contracted with any index in the case of rectangular Young diagrams.

There is actually a second conformally invariant condition that can be imposed on
mixed-symmetry tensors. This was found for hook diagrams in \cite{Alkalaev:2012rg} and requires a value for $\Delta$
different from the critical dimension.
It is now very easy to find the dimension where this condition can be imposed for general mixed-symmetry tensors
simply by lifting the conservation condition to the embedding space. This is most easily seen in the symmetric basis,
so now take $f$ to be in the symmetric basis as in \eqref{sym_basis_explicit} and consider the conservation condition
\beq
\frac{\partial}{\partial x_{g_1}} f_{(a_1\ldots a_{l_1})(b_1\ldots b_{l_2})\ldots(g_1\ldots g_{l_{h_1}})} (x) = 0 \,.
\label{eq:conservation_sym}
\eeq
The lift to embedding space (\ref{eq:conserved_derivation_1}--\ref{eq:def_r_tensor}) works exactly as before.
Now \eqref{eq:symmetrization_ex} is used to bring the last bracket in \eqref{eq:def_r_tensor} into a form analogous to 
\eqref{eq:R_last_index},
\beq
-\big(  (h_1-1) - (l_{h_1}-1) \big) F_{A_1 \ldots A_{|\lambda|}}\,.
\eeq
The conservation condition \eqref{eq:conservation_sym} becomes
\beq
0= 
\!
\left[ \frac{\partial}{\partial P_{A_{|\lambda|}}} - \frac{1}{P \cdot \bar P}\! \left(\bar P \cdot \frac{\partial}{\partial P} \right)\!P^{A_{|\lambda|}} 
-\big(l_{h_1}-h_1  +d-1-\Delta\big) \frac{\bar P^{A_{|\lambda|}}}{P \cdot \bar P} \right] 
\!
F_{A_1 \ldots A_{|\lambda|}} (P) \,.
\label{eq:R_last_index_sym}
\eeq
This is conformally invariant for
\beq
\Delta = l_{h_1}-h_1  +d-1 \,.
\eeq
For rectangular Young diagrams, where $h_{l_1} = h_1$ and $l_{h_1}=l_1$ this is again the critical dimension.
However, in general, we have $h_{l_1} \leq h_1$ and $l_{h_1} \leq l_1$, hence the unitarity bound \eqref{eq:unitarity_bound} is
violated for non-rectangular diagrams and the operators for which \eqref{eq:R_last_index_sym}
is conformally invariant are non-unitary.

\section{Conformal blocks}
\label{sec:conformal_blocks}

In this section we shall show how the above methods can be used to compute the conformal blocks 
for arbitrary irreducible tensor representations of the conformal group. The basic idea is that a conformal block in the channel 
${\cal O}_1\,{\cal O}_2 \to {\cal O}_3\,{\cal O}_4$ can be written as a conformal integral of the
product of the 3-point function of the operators ${\cal O}_1$, ${\cal O}_2$ and the exchanged operator ${\cal O}$ of dimension
$\Delta$, times the 3-point function of the operators ${\cal O}_3$, ${\cal O}_4$ and the shadow of the exchanged operator $\tilde{\cal O}$ of dimension
$d-\Delta$ \cite{Dobrev:1975ru,Dolan:2000ut,Dolan:2011dv}. This method makes use of the shadow formalism of 
\cite{Ferrara:1972xe,Ferrara:1972ay,Ferrara:1972uq,Ferrara:1973vz}. In practice, however, 
one needs to remove from the final expression the contribution of the shadow operator  exchange  to the conformal block, 
which has the wrong OPE limit. 
This can be done rather efficiently by doing a monodromy projection of the above conformal integral, as proposed in \cite{SimmonsDuffin:2012uy}.\footnote{Such split of the 
operator and its shadow exchanges can also be done using the Mellin space representation of the conformal partial wave \cite{Costa:2012cb}.}

Conformal blocks are known for many cases involving external scalar operators and the exchange of spin $l$ symmetric tensors. These results can be reused for correlators of external 
spin $l$ operators by acting with differential operators on the conformal blocks for external scalars \cite{Costa:2011dw}, but new exchanged tensor representations can not be taken care of in this way.
Here we will follow closely the approach detailed in \cite{SimmonsDuffin:2012uy} to compute the conformal blocks, and show with a non-trivial example
 that the embedding methods here presented  can be used to compute conformal blocks with external and exchanged operators in arbitrary tensor representations of the conformal group.

The idea is to define a projector to the conformal multiplet of a given operator
which, when inserted into a four-point function, produces the conformal partial wave for the exchange of
that operator (and its descendants).
For an operator $\calO$ with conformal dimension $\Delta$ this projector has the form
\beq
| \calO | = \frac{1}{\calN_{\calO}} \int D^d P_0 D^d P_5 \, \big| \calO \big(P_0; \bD_{\bZ_0}\big) \ket
\left. \bra \calO \big(P_0; \bZ_0\big) \calO \big(P_5; \bD_{\bZ_5}\big)\ket \right|_{\Delta \to \tilde \Delta}
\bra \calO (P_5; \bZ_5)\big| \,.
\label{eq:shadow_projector1}
\eeq
Note that we are schematically representing the index contraction of ${\cal O}$ with a differential operator acting on the
polarization vectors, as explained in (\ref{eq:contraction_embedding}).  
The integrals appearing here are called conformal integrals and are defined as
\beq
\int D^d P = \frac{1}{\text{Vol} \ GL(1,\mathbb{R})^+} \int_{P^+ +P^- \geq 0}
\text{d}^{d+2} P \  \delta(P^2) \,.
\eeq
Explicit expressions for these integrals are known for all functions that
appear in the computation of conformal blocks (see for instance appendix A.5 in \cite{Goncalves:2014rfa}).

The projector \eqref{eq:shadow_projector1} can be more compactly expressed in terms
of the shadow operator $\tilde \calO$, which is in the same $SO(d)$ irrep as $\calO$
and has conformal dimension $\tilde \Delta = d - \Delta$,
\beq
|\calO| = \frac{1}{\calN_{\calO}} \int D^d P_0  \, \big| \calO \big(P_0; \bD_{\bZ_0}\big) \ket
\bra \tilde \calO \big(P_0; \bZ_0\big)\big| \,,
\label{eq:shadow_projector2}
\eeq
where
\beq
\bra \tilde \calO (P_0; \bZ_0)\big|
=
 \int D^d P_5 
\left. \bra \calO \big(P_0; \bZ_0\big) \calO \big(P_5; \bD_{\bZ_5}\big)\ket \right|_{\Delta \to \tilde \Delta}
\bra \calO \big(P_5; \bZ_5\big)\big| \,.
\label{eq:shadow}
\eeq
Consider for simplicity the case where the three-point functions have only one tensor structure.
Inserting $|\calO|$ into a four-point function one obtains the 
conformal partial wave
\begin{align}
W_\calO ={}& \bra \calO_1 \big(P_1; \bZ_1\big) \calO_2 \big(P_2; \bZ_2\big) \big| \calO \big|
\calO_3 \big(P_3; \bZ_3\big) \calO_4 \big(P_4; \bZ_4\big) \ket 
\label{eq:conformal_partial_wave_from_proj}
\\
={}& \frac{1}{\calN_{\calO}} \int D^d P_0 \,
 \bra \calO_1 \big(P_1; \bZ_1\big) \calO_2 \big(P_2; \bZ_2\big) \calO \big(P_0; \bD_{\bZ_0}\big) \ket
\bra \tilde \calO \big(P_0; \bZ_0\big)\calO_3 \big(P_3; \bZ_3\big) \calO_4 \big(P_4; \bZ_4\big) \ket \,.
\nonumber
\end{align}
Since $\tilde \calO$ is in the same $SO(d)$ irrep as $\calO$, three-point functions
containing either of them must be equal, up to an overall constant and to the conformal dimensions of the operators, i.e.
\beq
\bra \tilde \calO \big(P_0; \bZ_0\big)\calO_3 \big(P_3; \bZ_3\big) \calO_4 \big(P_4; \bZ_4\big) \ket 
= \calS_\Delta \left. \bra \calO \big(P_0; \bZ_0\big)\calO_3 \big(P_3; \bZ_3\big) \calO_4 \big(P_4; \bZ_4\big)
\ket \right|_{\Delta \to \tilde \Delta} \,.
\label{eq:shadow_3pt_constant_def}
\eeq
This constant $\calS_\Delta$ is calculated by using the definition of the shadow operator \eqref{eq:shadow} and by computing the corresponding conformal integral.
The constant $\calN_{\calO}$ in \eqref{eq:conformal_partial_wave_from_proj} can then be calculated by demanding that $|\calO|$ acts trivially when
inserted into a three-point function, i.e.\ requiring 
\beq
\bra  \calO \big(P_0; \bZ_0\big)\big| \calO\big|  \calO_3 \big(P_3; \bZ_3\big) \calO_4 \big(P_4; \bZ_4\big) \ket
=
\bra  \calO \big(P_0; \bZ_0\big)  \calO_3 \big(P_3; \bZ_3\big) \calO_4 \big(P_4; \bZ_4\big) \ket \,.
\eeq
Using \eqref{eq:shadow_projector2} and \eqref{eq:shadow} one sees that this insertion amounts to
doing the shadow transformation twice, hence with \eqref{eq:shadow_3pt_constant_def} we have
\bea
 \bra  \calO \big(P_0; \bZ_0\big)| \calO|  \calO_3 \big(P_3; \bZ_3\big) \calO_4 \big(P_4; \bZ_4\big) \ket
={}& \frac{1}{\calN_{\calO}} \, \bra \tilde {\tilde \calO} \big(P_0; \bZ_0\big)  \calO_3 \big(P_3; \bZ_3\big) \calO_4 \big(P_4; \bZ_4\big) \ket \\
={}& \frac{\calS_\Delta \calS_{\tilde \Delta}}{\calN_{\calO}}\,
\bra  \calO \big(P_0; \bZ_0\big)  \calO_3 \big(P_3; \bZ_3\big) \calO_4 \big(P_4; \bZ_4\big) \ket \,,
\eea{eq:calN_computation}
and thus $\calN_{\calO} = \calS_\Delta \calS_{\tilde \Delta}$.

\subsection{Example: Hook diagram exchange}

As an example we will compute the conformal block $g_T^{\Delta_i}(u,v)$ for the exchange of 
the tensor $T$ with irreducible representation $[\,\bts{young_11},\Delta]$
in the correlation function of two scalars and two vectors $\langle \phi_1 J_2^\mu \phi_3 J_4^\nu \rangle$.
The conformal partial wave is
\begin{align}
&W_T =
\left( \frac{P_{14}}{P_{13}} \right)^\frac{\Delta_{34}}{2}
\left( \frac{P_{24}}{P_{14}} \right)^\frac{\Delta_{12}}{2}
\frac{g_T^{\Delta_i}(u,v)}{
P_{12}^{\frac{\Delta_1+\Delta_2}{2}} P_{34}^{\frac{\Delta_3+\Delta_4}{2}}}
\nonumber\\
={}&\bra \phi_1 \big(P_1\big) J_2 \big(P_2; Z_2\big) \big| T \big|
\phi_3 \big(P_3\big) J_4 \big(P_4; Z_4\big) \ket 
\label{eq:conformal_partial_wave_example}\\
={}& \frac{1}{\calS_{\tilde \Delta}} \int D^d P_0 
\left. \bra \phi_1 \big(P_1\big) J_2 \big(P_2; Z_2\big) T \big(P_0 ; D_{Z_0}, \partial_{\Theta_0}\big) \ket
\bra T \big(P_0 ; Z_0, \Theta_0\big) \phi_3 \big(P_3\big) J_4 \big(P_4; Z_4\big) \ket \right|_{\Delta \to \tilde \Delta}\,,
\nonumber
\end{align}
where we recall that $u,v$ are the cross ratios defined in (\ref{eq:cross_ratios}) and that the function $g_T^{\Delta_i}(u,v)$ also depends
on the external polarization vectors $Z_2$ and $Z_4$.
The ingredients for this calculation are the two- and three-point functions from \eqref{eq:ex_hook_2point}
and \eqref{eq:3pt_hook_scalar_vector}, for which we choose the normalizations
\bea
\bra T \big(P_1; Z_1, \Theta_1\big) \,T \big(P_2; Z_2, \Theta_2\big) \ket
={}& \frac{
2 \left(
H_{12}^{(\Theta,\Theta)}
 H_{12}^{(Z,Z)} 
- H_{12}^{(\Theta,Z)} H_{12}^{(Z,\Theta)} \right)
H_{12}^{(Z,Z)}
}{(P_{12})^{\Delta+3}} \,,\\
\!\!\!
\bra T \big(P_0 ; Z_0, \Theta_0\big) \phi_3 \big(P_3\big) J_4 \big(P_4; Z_4\big) \ket
={}& \frac{V_{0,34}^{(\Theta)} V_{0,34}^{(Z)} H_{04}^{(Z,Z)} - \left( V_{0,34}^{(Z)}\right)^2 H_{04}^{(\Theta,Z)}}
{(P_{03})^{\frac{\Delta+\Delta_3-\Delta_4+2}{2}} (P_{34})^{\frac{\Delta_3+\Delta_4-\Delta-2}{2}} (P_{40})^{\frac{\Delta_4+\Delta-\Delta_3+4}{2}}} \,,
\eea{eq:hook_normalizations}
the differential operator $D_Z$ from \eqref{eq:todorov_op_young11} which encodes the projection to the irrep $\bts{young_11}$,
the constant $\calS_{\Delta}$ and the solution of the conformal integrals.

The constant $\calS_{\Delta}$ is computed using \eqref{eq:shadow_3pt_constant_def}
and evaluating the conformal integral
\bea
&\bra \tilde T \big(P_0; Z_0, \Theta_0\big) \phi_3 \big(P_3\big) J_4 \big(P_4; Z_4\big) \ket \\
={}&
\int D^d P_5 \left. \bra T \big(P_0; Z_0, \Theta_0\big) \,T \big(P_5; D_{Z_5}, \partial_{\Theta_5}\big) \ket \right|_{\Delta \to \tilde \Delta}
\bra  T \big(P_5; Z_5, \Theta_5\big) \phi_3 \big(P_3\big) J_4 \big(P_4; Z_4\big) \ket \,.
\eea{eq:hook_compute_calS}
All the integrals here are of the type
\beq
\int D^d P_5 \,\frac{P_5^{A_1}\ldots P_5^{A_n}}{(P_{50})^a (P_{53})^b (P_{54})^c} \,,
\eeq
and their explicit solution can be found for instance in \cite{Cornalba:2008qf,Goncalves:2014rfa}.\footnote{
To give an impression of how these integrals look like, here is the case with $n=1$
\begin{align}
\int D^d P_5 \frac{P_5^{A}}{(P_{50})^a (P_{53})^b (P_{54})^c} 
={}&\frac{ \Gamma\!\left( \frac{b+c-a+1}{2} \right) \Gamma\!\left( \frac{c+a-b+1}{2}   \right)  \Gamma\!\left( \frac{a+b-c+1}{2} \right)}
{\Gamma(a) \,\Gamma(b) \,\Gamma(c)  }
\frac{\pi^h}{(P_{34})^{\frac{b+c-a+1}{2}} (P_{40})^{\frac{c+a-b+1}{2}} (P_{03})^{\frac{a+b-c+1}{2}} }
\nonumber\\
&\times\left( 
\frac{P_{34}P_0^A}{\frac{1}{2} (b+c-a+1)}
+\frac{P_{40}P_3^A}{\frac{1}{2} (c+a-b+1)}
+\frac{P_{03}P_4^A}{\frac{1}{2} (a+b-c+1)}
\right).
\label{eq:3pt_vector_integral}
\end{align}}
Comparing the integral in  \eqref{eq:hook_compute_calS} with the three-point function, the resulting constant is
\beq
\calS^{\bts{young_11}}_\Delta = \frac{\pi^h (\Delta-2) \Delta \,\Gamma(\Delta-h)}{\Gamma(\tilde \Delta + 2)}
\frac{
\Gamma\!\left( \frac{\tilde \Delta + \Delta_{34} + 2}{2} \right)
\Gamma\!\left( \frac{\tilde \Delta - \Delta_{34} + 2}{2} \right)
}{
\Gamma\!\left( \frac{\Delta + \Delta_{34} + 2}{2} \right)
\Gamma\!\left( \frac{ \Delta - \Delta_{34} + 2}{2} \right)
} \,.
\eeq
Note that this is very similar to the corresponding constant for the exchange of the 
antisymmetric two-tensor $\bts{young_01}\,$, given below in \eqref{eq:S_twoform},
which was calculated in \cite{SimmonsDuffin:2012uy}. As a small consistency check observe
that the constant $\calN_{\calO} = \calS_\Delta \calS_{\tilde \Delta}$
appearing in \eqref{eq:shadow_projector1} is independent of $\Delta_{34}$.

To calculate the conformal partial wave \eqref{eq:conformal_partial_wave_example}
it is enough to know the conformal integrals 
\beg
\int D^d P_0 \,\frac{(P_0 \cdot Z_2) (P_0 \cdot Z_4)}{(P_{01})^a (P_{02})^b (P_{03})^e (P_{04})^f} \,, \qquad
\int D^d P_0 \,\frac{P_0 \cdot Z_2}{(P_{01})^a (P_{02})^b (P_{03})^e (P_{04})^f} \,,\\
\int D^d P_0 \,\frac{P_0 \cdot Z_4}{(P_{01})^a (P_{02})^b (P_{03})^e (P_{04})^f}\,, \qquad
\int D^d P_0 \,\frac{1}{(P_{01})^a (P_{02})^b (P_{03})^e (P_{04})^f} \,,
\eeg{eq:4pt_conf_intetgrals}
which much like the example \eqref{eq:3pt_vector_integral} can be brought into a form where the polarizations are contracted with $P_1, P_2,P_3$ and $P_4$, or with each other.
Just as in \cite{SimmonsDuffin:2012uy}, after doing the monodromy projection to eliminate the shadow block, 
the final expression depends on functions of the cross ratios $u,v$ given by 
\bea
J_{j,k,l}^{(i)} = \frac{\Gamma(h+i-f) \,\Gamma(f) \sin (\pi f)}{\sin \!\big(\pi(e+f-h-i)\big)} &
\int\limits_0^\infty \frac{\dd x}{x} \int\limits_{x+1}^\infty \frac{\dd y}{y} \frac{x^b y^e}{(y+v x y-ux)^{h+i-f} (y-x-1)^f} \,, 
\eea{eq:J_def}
with
\bea
b &= \alpha +i+j-1 \,, \\
e &= \beta-\Delta +h +i+k-l\,,\\
f &= 1-\beta +h-k \,,
\eea{eq:J_parameters_def}
and
\beq
\alpha = \frac{\Delta-\Delta_{12}-2}{2} \,, \qquad \beta = \frac{\Delta+\Delta_{34}-2}{2} \,,
\eeq
where $\Delta_{ij}=\Delta_{i}-\Delta_{j}$ and $h=d/2$.
In even dimensions, $h \in \mathbb{N}$, the functions $J_{j,k,l}^{(i)}$ can be expressed in terms of ${}_2{F}_1$
hypergeometric functions, see \cite{SimmonsDuffin:2012uy}.

Doing the computation we arrived at the following expression for the conformal block defined in (\ref{eq:conformal_partial_wave_example}),
\begin{align}
\nonumber
&g^{\De_i}_T(u,v)=\frac{u^{\De/2-1}\G(\De+2)}
{4P_{24}(\tilde\De-2)\tilde\De(2h-1)\,\G(\a +2)\,\G(\b +2)\,\G(\De-\a )\,\G(\De-\b )\,\G(h-\De)}
\\
&
\times\left[
V^{(Z)}_{2,14} V^{(Z)}_{4,12} \,u F_1
+
V^{(Z)}_{2,14} V^{(Z)}_{4,23} \,v F_2
+
V^{(Z)}_{2,34} V^{(Z)}_{4,12} \,u F_3
+
V^{(Z)}_{2,34} V^{(Z)}_{4,23} \,v F_4
+
\frac{1}{2} H^{(Z,Z)}_{24}
F_H
\right].
\label{eq:hook_conformal_block}
\end{align}
As expected, this conformal block is organized into tensor structures that are analogous to the ones
discussed for this four-point correlator in Section \ref{sec:example_sc-v-sc-v}.
The functions $F_i=F_i(u,v)$  depend on $h$, $\Delta$, $\alpha$ and $\beta$, and are expressed in terms of a finite number of the
integrals $J_{j,k,l}^{(i)}$ given in (\ref{eq:J_def}) above. For clarity of exposition we decided to present these functions in the Appendix \ref{app:conformal_block}.\footnote{
A Mathematica notebook containing this result can be obtained from the authors upon request.}

The example at hand shows that we have a well defined algorithm to compute any conformal block. However, 
before going on to compute even more complicated conformal blocks, it would be helpful to study the functions
$J_{j,k,l}^{(i)}$ in detail. Once the relations among them are better understood, it may well be that much
shorter expressions for the conformal blocks are possible. We hope to return to this question.

\subsection{Example: Two-form exchange}

The conformal block for exchange of a two-form tensor $F$, which corresponds to the irrep $\bts{young_01}\,$,
was computed analogously in \cite{SimmonsDuffin:2012uy},
however the result contained a few typos which we now correct.\footnote{We thank David Simmons-Duffin for correspondence on this point.} 
The normalizations for the two- and three-point functions of \cite{SimmonsDuffin:2012uy} are in our notation
\bea
\bra F \big(P_1, \Theta_1\big) F \big(P_2,  \Theta_2\big) \ket
={}& 
\frac{1}{4}
\frac{
\left(
H_{12}^{(\Theta,\Theta)}
 \right)^2
}{(P_{12})^{\Delta+2}} \,,\\
\bra F \big(P_0 ,  \Theta_0\big) \phi_3 \big(P_3\big) J_4 \big(P_4, Z_4\big) \ket
={}& 
\frac{V_{0,34}^{(\Theta)}  H_{04}^{(\Theta,Z)} }
{(P_{03})^{\frac{\Delta+\Delta_3-\Delta_4+1}{2}} (P_{34})^{\frac{\Delta_3+\Delta_4-\Delta-1}{2}} (P_{40})^{\frac{\Delta_4+\Delta-\Delta_3+3}{2}}} \,,
\eea{eq:2form_normalizations}
and the contraction of two-forms is now done using the normalized derivative $\partial_\Theta/\sqrt{2}$.
The constant $\calS_\Delta$ is given by
\beq
\calS^\bts{young_01}_\Delta = \frac{\pi^h (\Delta-2) \,\Gamma(\Delta-h)}{4\,\Gamma(\tilde \Delta + 1)}
\frac{
\Gamma\!\left( \frac{\tilde \Delta + \Delta_{34} + 1}{2} \right)
\Gamma\!\left( \frac{\tilde \Delta - \Delta_{34} + 1}{2} \right)
}{
\Gamma\!\left( \frac{\Delta + \Delta_{34} + 1}{2} \right)
\Gamma\!\left( \frac{ \Delta - \Delta_{34} + 1}{2} \right)
} \,.
\label{eq:S_twoform}
\eeq
After doing carefully the conformal integrals we obtained a slightly shorter formula for this conformal block,
\bea
&g^{\De_i}_F(u,v)=\frac{2 u^{\De/2-1/2} \,\G(\De+1)}{P_{24}(2-\tilde\De)\,\G(\a +1)\,\G(\b +1)\,\G(\De-\a )\,\G(\De-\b )\,\G(h-\De)}
\\
&
\times\Bigg[V^{(Z)}_{2,14} V^{(Z)}_{4,12} \,u
\Bigl( 
(v-1) J^{(2)}_{0,1,2}+v (\b-h+1) J^{(1)}_{1,2,2}+(\b-\De+h)J^{(1)}_{1,1,2}
\Bigr)
\\
&\quad
-
V^{(Z)}_{2,14} V^{(Z)}_{4,23} \,v (v-1) J^{(2)}_{0,1,1}
\\
&\quad
+
V^{(Z)}_{2,34} V^{(Z)}_{4,12} \,u \biggl(
(v-1) J^{(2)}_{0,1,1}- \a J^{(1)}_{0,1,1} 
+(\b-h+1) \left(v J^{(1)}_{1,2,1} - \a  J^{(0)}_{1,2,1}\right)
\\
&\quad\quad\quad\quad\quad\quad\quad\quad
+(\b-\a+h-1)J^{(1)}_{1,1,1}
\biggr)
\\
&\quad
+
V^{(Z)}_{2,34} V^{(Z)}_{4,23}\,
v
\Bigl(
-(v-1)J^{(2)}_{0,1,0}
+\a  J^{(1)}_{0,1,0}
+(\a -\Delta+1 ) J^{(1)}_{1,1,0}
\Bigr)
\\
&\quad
-
\frac{ 1}{2} H^{(Z,Z)}_{24}
\biggl(
-\frac{v-1}{h+1} \Big(J^{(2)}_{0,0,0} + J^{(2)}_{0,0,1} + J^{(2)}_{0,1,1} +
J^{(2)}_{1,0,1} + v J^{(2)}_{1,1,1} + u J^{(2)}_{1,1,2}\Big)
\\
&
\quad\quad\quad\quad\quad\quad\quad
+(\b -\Delta+h ) \Big(\a J^{(0)}_{1,1,1}+(\a -\Delta+1 )J^{(0)}_{2,1,1}\Big)
\\
&
\quad\quad\quad\quad\quad\quad\quad
+(\b -h+1) \Big(\a J^{(0)}_{1,2,1}+(\a -\Delta+1 ) v J^{(0)}_{2,2,1}\Big)
\biggr)
\Bigg]\,,
\eea{eq:cb_2form}
where $J_{j,k,l}^{(i)}$ is defined in \eqref{eq:J_def}, but now with
\beq
\alpha = \frac{\Delta-\Delta_{12}-1}{2} \,, \qquad \beta = \frac{\Delta+\Delta_{34}-1}{2} \,.
\eeq
To compare this to the corrected result of \cite{SimmonsDuffin:2012uy}, we took into account the different definitions for $H_{ij}$ and $V_{i,jk}$, and 
used the three following identities which we checked numerically,
\begin{align}
u&
\Bigl( 
(v-1) J^{(2)}_{0,1,2}+v (\b-h+1) J^{(1)}_{1,2,2}+(\b-\De+h)J^{(1)}_{1,1,2}
\Bigr)
\nonumber\\
&\ =
(\beta -\Delta +h) \left( \alpha  J_{1,1,1}^{(0)} -J_{1,0,1}^{(1)}\right) 
-v(\beta -h+1) \left(J_{2,2,1}^{(0)} (\alpha -\Delta +1) + J_{1,2,1}^{(1)}\right) 
\label{eq:J_id1}
\\
&\ -\alpha  J_{0,1,1}^{(1)}-2 v (\alpha +\beta -\Delta +1)J_{1,1,1}^{(1)} -v (\alpha -\Delta +1) \left(J_{1,1,0}^{(1)} +   J_{2,1,1}^{(1)} \right)
\nonumber\\
&\ -\alpha  v J_{0,1,0}^{(1)}+(v-1) \left(v J_{0,1,0}^{(2)}+ v J_{1,1,1}^{(2)} - J_{0,0,1}^{(2)}\right) ,
\nonumber
\\
-&u J^{(2)}_{0,1,1}
= 
J^{(2)}_{0,0,0}+ v J^{(2)}_{0,1,0}-\a  J^{(1)}_{0,1,0} \,,
\\
\left( v\right.&\left.-1\right) J^{(2)}_{0,1,1}- \a J^{(1)}_{0,1,1} 
+(\b-h+1) \left(v J^{(1)}_{1,2,1} - \a  J^{(0)}_{1,2,1}\right)
+(\b-\a+h-1)J^{(1)}_{1,1,1}
\nonumber\\
&\ =
-(v-1)\left(J^{(2)}_{0,1,0}+J^{(2)}_{1,1,1}\right)
+(\b -h+1)\left(J^{(1)}_{1,2,1}+(\a -\Delta+1 ) J^{(0)}_{2,2,1}\right)
+\a  J^{(1)}_{0,1,0}
\label{eq:compare_cb_terms}
\\
&\ 
+(\a +\b -\Delta+h ) J^{(1)}_{1,1,1}
+(\a -\Delta+1 ) \left(J^{(1)}_{1,1,0}+J^{(1)}_{2,1,1}\right).
\nonumber
\end{align}

\section{S-matrix rule for counting structures}

The matching of tensor structures in CFT correlators and scattering amplitudes that was found for
symmetric tensors in \cite{Costa:2011mg} straightforwardly generalizes to general irreps when
considering non-conserved operators. The general statement is:
\textit{
The number of independent structures in a correlation function of n non-conserved operators of $SO(d)$
irreps ${\lambda_1, \ldots, \lambda_n}$ is equal to to the number of independent structures in a $n$-point scattering
amplitude of massive particles of the same irreps in $d+1$ dimensional flat Minkowski space.
}

This is not surprising since particles have polarizations in irreps of the little group,
which is $SO(d)$ for massive particles in $d+1$ dimensions. 
The index-free notation introduced in Section \ref{sec:index_free_notation} can be employed by simply using
the same Young-symmetrized polarizations.
Thus,
an $n$-point scattering amplitude of irreps $\lambda_1, \ldots, \lambda_n$ and momenta $k_1, \ldots, k_n$ can be written as
\bea
A_{\lambda_1 \ldots \lambda_n} (\{k_i;\bz_i\})
=
\prod\limits_{j=1}^{n}
\prod\limits_{p=1}^{\nz^j} \prod\limits_{q=1}^{\min ( l_p^j,\ntheta^j)}
\left( z^{(p)}_j \cdot \partial_{\theta^{(q)}_j} \right) 
\sum\limits_k f_k(v_a) \,\bar R_{\lambda_1 \ldots \lambda_n}^{(k)} \big(\{k_i;\btheta_i\}\big) \,,
\eea{eq:n-point_amplitude}
where $f_k(v_a)$ are functions of the $n(n-3)/2$ independent Mandelstams $v_a$.
The momenta and polarizations are vectors in $(d+1)$-dimensional Minkowski space
and the polarizations are transverse to the corresponding momenta
\beq
\theta_i^{(p)} \cdot k_i = z_i^{(p)} \cdot k_i = 0\,.
\label{eq:amplitude_transversality}
\eeq
The scaling in the polarization vectors is fixed
by the condition that the complete polarization tensor appears linearly in the amplitude.
This translates to the following scaling of $\bar R_{\lambda_1 \ldots \lambda_n}^{(k)}$ in the polarization vectors,
which is equivalent to the one  for CFT correlators  in \eqref{eq:npoint-scale},
\bea
\bar R_{\lambda_1 \ldots \lambda_n}^{(k)}
\big(\{k_i; \bbeta_{i} \btheta_i\}\big)
= \bar R_{\lambda_1 \ldots \lambda_n}^{(k)}
\big(\{k_i;\btheta_i\}\big) \prod\limits_i \left(\beta_i^{(1)}\right)^{h^i_1} \ldots \left(\beta_i^{(\ntheta^i)} \right)^{h^i_{\ntheta^i}} \left( \beta_i^{(Z)}\right)^{(\lambda_i)_1} .
\eea{eq:amplitude-scale}
The functions $\bar R_{\lambda_1 \ldots \lambda_n}^{(k)}$ can be constructed from the two kinds of building blocks
\beq
\tilde H^{(p,q)}_{ij} \equiv \theta_i^{(p)} \cdot \theta_j^{(q)} \,, \quad\quad
\mathcal{\tilde V}_{ij}^{(p)} \equiv \theta_i^{(p)} \cdot k_j \,,
\label{eq:amplitude_building_blocks}
\eeq
where $\theta_i^{(p)}$ should be replaced by $z_i^{(1)}$ for $p=z$.
There are $n-2$ independent $\mathcal{\tilde V}_{ij}^{(p)}$'s for each $i$, because one of the possible terms vanishes due to the transversality condition
\eqref{eq:amplitude_transversality} and another one can be eliminated using momentum conservation
\beq
k_1 + k_2 +\ldots + k_n = 0\,.
\eeq
Furthermore, the building blocks depend in the same way on Grassmann polarizations as their counterparts $H^{(p,q)}_{ij}$ and $\mathcal{V}_{ij}^{(p)} $ that appear in CFT correlators.
Hence, there is a one-to-one correspondence between building blocks and the counting of tensor structures is the same as in CFT correlators.

A more thorough treatment of on-shell amplitudes of arbitrary $SO(d)$ irreps 
(in the context of the open bosonic string) can be found in \cite{Boels:2014dka}.

\section{Concluding remarks}

In this work we developed a formalism to elegantly describe irreducible tensor representations of $SO(d)$
in terms of polynomials. With this formalism and the help of representation theory,
tensor structures in CFT correlators and scattering amplitudes become tangible.
We gave an algorithm for counting the number of independent tensor structures
in any CFT correlator (or massive scattering amplitude) of bosonic operators (or particles),
allowing for a systematic construction of the tensor structures
for any given example.

The most obvious application for correlators of mixed-symmetry tensors is the
construction of conformal blocks, which we reviewed using our new index-free notation.
Once all conformal blocks appearing in a given
correlator are known, it is possible to implement constrains that follow from conformal symmetry, using recent 
conformal bootstrap techniques, i.e.\ 
proving bounds on the CFT data
(conformal dimensions $\Delta_i$ and OPE coefficients) by use of linear programming \cite{Rattazzi:2008pe}. Since there are no further
assumptions, such bounds are universal, they hold for any CFT.
Until now, in lack of conformal blocks for mixed-symmetry tensor exchange, this has only be done for correlators of scalar operators.

While we only computed one conformal block of mixed-symmetry tensor
exchange in a correlator of two scalars and two vectors,
it would be much more interesting to consider correlators of stress-tensors.
This is because the stress-tensor appears in any CFT and thus could lead to truly
universal bounds on CFT data. Another reason for interest in the stress-tensor
is its connection to the graviton in AdS, via the AdS/CFT duality. As was pointed out already in \cite{SimmonsDuffin:2012uy},
universal bounds on CFT data for external operators with spin may explain the weak gravity conjecture
\cite{ArkaniHamed:2006dz} or the bounds on $a$ and $c$ in \cite{Hofman:2008ar}.

\begin{table}[t!]
\centering
{\renewcommand{\arraystretch}{2}
\begin{tabular}{ c  c  }
correlator & new exchanged $SO(d)$ irreps \\ \hline
$\langle \phi_1  \phi_2 \phi_3 \phi_4 \rangle$ & $\btm{young_hook0}$ \\  
$\langle \phi_1  J_2^\mu \phi_3 J_4^\nu \rangle$ & $\btm{young_hook1}$ \\  
$\langle J_1^\mu  J_2^\nu J_3^\rho J_4^\sigma \rangle$ & $\btm{young_hook2a}$, \ $\btm{young_hook2b}$\\  
$\langle J_1^\mu  T_2^{\nu \rho} J_3^\sigma T_4^{\lambda \kappa} \rangle$ & $\btm{young_hook3a}$, \ $\btm{young_hook3b}$\\  
$\langle T_1^{\mu \nu}  T_2^{\rho \sigma} T_3^{\lambda \kappa} T_4^{\tau \omega} \rangle$ & $\btm{young_hook4a}$, \ $\btm{young_hook4b}$, \ $\btm{young_hook4c}$ \\ 
\end{tabular}
}
\caption[]{Exchanged irreps in correlators of currents and stress-tensors,
following the discussion of possible tensor structures 
for three-point functions in Section \ref{sec:three-point_functions} and the construction of conformal blocks in
Section \ref{sec:conformal_blocks}.}
\label{table:blocks_in_correlators}
\end{table}

With the insights about three-point correlators from this work
it is easy to outline what needs to be done to compute all conformal blocks
for the correlator of four stress-tensors.
Table \ref{table:blocks_in_correlators} contains all irreps that are exchanged in this correlator.
Some conformal blocks can actually be written in terms of derivatives
of conformal blocks for exchange of the same irrep in a simpler correlator,
as it is the case for exchange of symmetric tensors \cite{Costa:2011mg}. 
For example,
the conformal blocks for exchange of $\bts{young_hook0}$ in 
$\langle T_1^{\mu \nu}  T_2^{\rho \sigma} T_3^{\lambda \kappa} T_4^{\tau \omega} \rangle$
are given by derivatives of the conformal blocks of
$\langle \phi_1  \phi_2 \phi_3 \phi_4 \rangle$.
For this reason, 
each line in the table displays for some correlator the irreps of exchanged operators 
that appear for the first time for that correlator.
If one picks a correlator in one line and a single irrep from the same line,
the computation of that conformal block is comparatively easy, since in those cases the three-point function between external operators and the 
exchanged operator has only one tensor structure. 
One can hope that the conformal blocks for all other cases are given by derivatives of those simpler cases.

An interesting generalisation of our work would be  to extend the formalism to general spinor representations of $SO(d)$.
This would complete the counting and construction of tensor structures for all CFT correlators
and facilitate the conformal bootstrap for combinations of operators that imply exchange of operators with half-integer spin.

Finally, note that most discussions of higher spin fields in AdS focus on the case of spin $J$ symmetric tensors. However, it would be interesting to
consider AdS fields  dual to operators in arbitrary irreps of the conformal group. 
We expect that the techniques described in this paper can also be extended to the case of AdS fields, in the spirit
of \cite{Costa:2014kfa}.

\acknowledgments
We wish to thank Rutger Boels, Anders Skovsted Buch, Vasco Gon\c{c}alves, Christoph Horst, 
Jo\~ao Penedones and Emilio Trevisani  for helpful discussions. T.H.\ thanks Universidade do Porto for hospitality. 
The research leading to these results has received funding from the German Science Foundation (DFG) within the Collaborative Research Center 676 ``Particles, Strings and the Early Universe'',
from the [European Union] Seventh Framework Programme [FP7-People-2010-IRSES] and [FP7/2007-2013] under grant agreements No 269217 and 317089, and from the grant CERN/FP/123599/2011.
\emph{Centro de Fisica do Porto} is partially funded by the Foundation for 
Science and Technology of Portugal (FCT). 

\appendix  
\section{Functions in the conformal block for hook diagram exchange}
\label{app:conformal_block}

The following are the functions appearing in the conformal block \eqref{eq:hook_conformal_block} for exchange of
the primary in the irreducible representation $[\,\bts{young_11},\Delta]$
in the correlator of two scalars and two vectors  $\langle \phi_1 J_2^\mu \phi_3 J_4^\nu \rangle$.\footnote{
Many thanks to Fernando Rejon-Barrera for pointing out typos in previous versions of the expressions $F_4$ and $F_H$.}
\begin{align}
F_1={}&(\alpha -\Delta +1)
\Bigg[
(\beta -\Delta +h+1)
\Big(
-(2 h-1) J_{2,1,2}^{(1)} (\beta -\Delta +h)-(\alpha +1) J_{1,1,1}^{(1)}
\nonumber\\
&-J_{1,1,2}^{(2)} \big((2 h-1) (v-1)+u\big)+2 (2 h-1) v J_{2,2,2}^{(1)} (-\beta +h-1)
\Big)
\nonumber\\
&+(\beta -h+1)\, v
\bigg(
(2 h-1)\Big(v J_{2,3,2}^{(1)} (-\beta +h-2) -(v-1)J_{1,2,2}^{(2)} \Big)
+u J_{1,2,2}^{(2)}
\bigg)
\Bigg]
\nonumber\\
&+(\alpha +1) 
\Bigg[
(\beta -\Delta +h+1)
\Big(
2 \alpha  J_{1,2,1}^{(0)} (-\beta +h-1)+(1-2 h) J_{1,1,2}^{(1)} (\beta -\Delta +h)
\nonumber\\
&-J_{1,2,2}^{(1)} (-\beta +h-1) \big((1-2 h) (v+1)+2 u\big)
+J_{0,1,2}^{(2)}
   \big((1-2 h) (v-1)+u\big)-\alpha  J_{0,1,1}^{(1)}
\Big)
\nonumber
\\
&+(\beta -h+1)
\Big(
J_{0,2,2}^{(2)} \big((1-2 h) (v-1)-u\big)
\nonumber\\
&+(2 h-1) \,v J_{1,3,2}^{(1)} (-\beta +h-2)+\alpha  J_{0,2,1}^{(1)}
+v(\alpha -\Delta +1) J_{1,2,1}^{(1)}
\Big)
\Bigg]\\
F_2={}&(\alpha -\Delta +1)
\Bigg[
(\beta -\Delta +h+1)
\Big(
u J_{2,2,2}^{(1)} (-\beta +h-1)+J_{1,1,1}^{(2)} \big((2 h-1) (v-1)+u\big)
\Big)
\nonumber\\
&+(\beta -h+1) \,v
\bigg(
(-\beta +h-2) \Big((\alpha +1) J_{2,3,1}^{(0)}+u J_{2,3,2}^{(1)}\Big)
\nonumber\\
&-J_{1,2,1}^{(2)} \big((1-2 h) (v-1)+u\big)-(\alpha +1) J_{1,2,0}^{(1)}
\bigg)
\Bigg]
\nonumber
\end{align}
\begin{align}
&+(\alpha +1) 
\Bigg[
(\beta -\Delta +h+1)
\bigg(
(\beta -h+1)
\Big(u J_{1,2,2}^{(1)} -
J_{2,2,1}^{(0)} (\alpha -\Delta +1) -\alpha  J_{1,2,1}^{(0)} \Big)
\nonumber\\
&-	J_{0,1,1}^{(2)} \big((1-2 h) (v-1)+u\big)+J_{1,1,0}^{(1)} (\alpha -\Delta +1)+\alpha  J_{0,1,0}^{(1)}
\bigg)
\nonumber\\
&+(\beta -h+1)
\bigg(
-\alpha  \Big(J_{1,3,1}^{(0)} (\beta -h+2)+J_{0,2,0}^{(1)}\Big)+u J_{1,3,2}^{(1)} (\beta -h+2)
\nonumber\\
&+J_{0,2,1}^{(2)} \big((2 h-1) (v-1)+u\big)
\bigg)
\Bigg]
\label{eq:V214V423_simplified}\\
F_3={}&(\alpha -\Delta +1)
\Bigg[
(\beta -\Delta +h+1)
\bigg(
J_{1,1,1}^{(2)} \big((1-2 h) (v-1)-u\big)
\nonumber\\
&-(2 h-1) \Big((\alpha +1) J_{2,2,1}^{(0)} (-\beta +h-1)+J_{2,1,1}^{(1)} (-\alpha +\beta +h-2) \Big)
\bigg)
\nonumber
\\
&+(\beta -h+1) v
\bigg(
(2 h-1) (\beta -h+2) \Big( (\alpha +1) J_{2,3,1}^{(0)} -v J_{2,3,1}^{(1)} \Big)
\nonumber
\\
&
-(2 h-1) J_{2,2,1}^{(1)} (-\alpha +2 \beta -\Delta +2 h)  +J_{1,2,1}^{(2)} \big((1-2 h)  (v-1)+u\big)
\bigg)
\Bigg]
\nonumber
\\
&+(\alpha +1) 
\Bigg[
(\beta -\Delta +h+1)
\bigg(
J_{0,1,1}^{(2)} \big((1-2 h) (v-1)+u\big)
\nonumber\\
&
-(2 h-1) \Big(J_{1,1,1}^{(1)} (-2 \alpha +\beta +\Delta +h-2)-\alpha J_{1,2,1}^{(0)} (\beta -h+1)-\alpha J_{0,1,1}^{(1)}\Big)
\bigg)
\nonumber
\\
&+(\beta -h+1)
\bigg(
(2 h-1) (\beta -h+2)\Big(\alpha   J_{1,3,1}^{(0)} -v J_{1,3,1}^{(1)} \Big) +\alpha  (2 h-1) J_{0,2,1}^{(1)}
\nonumber
\\
&
+J_{1,2,1}^{(1)} ( \beta-\alpha +h) \big((1-2 h) (v+1)+2  u\big)
-J_{0,2,1}^{(2)} \big((2 h-1) (v-1)+u\big)
\bigg)
\Bigg]
\label{eq:V234V412_simplified}\\
F_4={}&(\alpha -\Delta +1)
\Bigg[
(\beta -\Delta +h+1)
\bigg(
(2 h-1) \Big( J_{2,1,0}^{(1)} (-\alpha +\Delta -2)-2 (\alpha +1)  J_{1,1,0}^{(1)} \Big)
\nonumber
\\
&+u J_{2,2,1}^{(1)} (-\beta +h-1)+J_{1,1,0}^{(2)} \big((2 h-1) (v-1)+u\big)
\bigg)
\nonumber
\\
&+(\beta -h+1)
\Big(
-v J_{1,2,0}^{(2)} \big((1-2 h) (v-1)+u\big)+(2 h-1) v J_{2,2,0}^{(1)} (-\alpha
   +\Delta -2)
\nonumber
\\
&+(\alpha +1) J_{1,2,0}^{(1)} \big((1-2 h) (v+1)+2 u\big)+u  (-\beta +h-2) \big( v J_{2,3,1}^{(1)} - 2 J_{2,3,1}^{(0)} (1+\alpha)\big)
\Big)
\Bigg]
\nonumber\\
&+(\alpha +1) 
\Bigg[
(\beta -\Delta +h+1)
\Big(
\alpha  (1-2 h) J_{0,1,0}^{(1)}+J_{0,1,0}^{(2)} \big((2 h-1) (v-1)-u\big)
\Big)
\nonumber\\
&+(\beta -h+1)
\Big(
\alpha  (1-2 h) J_{0,2,0}^{(1)}+u J_{1,3,1}^{(1)} (\beta -h+2)
\nonumber
\\
&+J_{0,2,0}^{(2)} \big((2 h-1) (v-1)+u\big)
+u J_{1,2,1}^{(1)} (\beta -\Delta +h+1)
\Big)
\Bigg]
\label{eq:V234V423_simplified}
\end{align}
\begin{align}
F_H={}&
\frac{1}{h+1}
\Bigg\{
(\alpha -\Delta +1)
\Bigg[
(\beta -\Delta +h+1)
\bigg(
  \big((1-2 h) (v-1)-u\big)
\nonumber\\
&
\times \Big( J_{1,1,1}^{(2)} + J_{1,0,0}^{(2)} + J_{2,0,1}^{(2)} +   u J_{2,1,2}^{(2)}\Big)
\bigg)
\nonumber\\
&+(\beta -h+1)
\bigg(
v \big((1-2 h) (v-1)+u\big) \Big(J_{1,1,1}^{(2)}+u J_{2,2,2}^{(2)}+v J_{2,2,1}^{(2)}+J_{1,1,0}^{(2)}\Big)
\bigg)
\nonumber\\
&+
v J_{2,1,1}^{(2)} \bigg((1-2 h) \Big(u-(v-1) \big(\Delta-2 (\beta+1)\big)\Big)+(\Delta-1)  u\bigg)
\Bigg]
\nonumber\\
&+(\alpha +1) 
\Bigg[
(\beta -\Delta +h+1)
\big(u-(2 h-1) (v-1)\big) \Big(J_{1,1,1}^{(2)} v
+J_{0,0,0}^{(2)}
+J_{0,0,1}^{(2)}
+u J_{1,1,2}^{(2)} \Big)
\nonumber\\
&+(\beta -h+1)
\big((1-2 h) (v-1)-u\big)
\Big(u J_{1,2,2}^{(2)}+J_{0,1,0}^{(2)}+J_{0,2,1}^{(2)}+J_{1,1,1}^{(2)}\Big) 
\nonumber\\
&+
J_{0,1,1}^{(2)} \bigg((2 h-1) \Big(u+(v-1) \big(\Delta -2 (\beta +1)\big)\Big)-(\Delta-1) u \bigg)
\Bigg]
\nonumber\\
&+
J_{1,0,1}^{(2)} (\beta -\Delta +h+1) \Big((2 h-1) (v-1) \big(\Delta-2 (\alpha +1)\big)+\Delta  u\Big)
\nonumber
\\
&-v J_{1,2,1}^{(2)}
   (-\beta +h-1) \Big((2 h-1) (v-1) \big(\Delta -2 (\alpha +1)\big)-\Delta u\Big)
\Bigg\}
\nonumber\\
&+
2 J_{2,2,1}^{(0)}(\alpha +1)  (-\alpha +\Delta -1) (-\beta +h-1) (\beta -\Delta +h+1) \big((2 h-1) (v+1)-2 u\big)
\nonumber\\
&+
(2h-1)
\Bigg\{
(\alpha -\Delta +1)
\Bigg[
(\beta -\Delta +h+1)
\bigg(
2 (\alpha +1) J_{2,1,1}^{(0)} (\beta -\Delta +h)
\nonumber\\
&
+(\alpha -\Delta +2) \Big(J_{3,1,1}^{(0)}  (\beta -\Delta +h) +2 v J_{3,2,1}^{(0)}  (\beta -h+1) \Big)
\bigg)
\nonumber\\
&+(\beta -h+1)
\bigg(
v (-\beta +h-2) \Big(v J_{3,3,1}^{(0)} (-\alpha +\Delta -2)-2 (\alpha +1) J_{2,3,1}^{(0)}\Big)
\bigg)
\Bigg]
\nonumber\\
&+\alpha (\alpha +1) 
\Bigg[
 (\beta -\Delta +h+1)   \Big(-2 J_{1,2,1}^{(0)} (-\beta +h-1) + J_{1,1,1}^{(0)} (\beta -\Delta +h)  \Big)
 \nonumber\\
& +J_{1,3,1}^{(0)}
   (-\beta +h-2) (-\beta +h-1)
\Bigg]
\Bigg\}
\label{eq:H24_simplified}
\end{align}
\newpage

\bibliographystyle{JHEP}

\bibliography{mixed}

\end{document}